\documentstyle[prl,aps,epsf,12pt]{revtex}
\begin{document}
\title{\large{\bf The Addition Spectrum and Koopmans' Theorem for
Disordered Quantum Dots}}
\author{Paul N. Walker$^1$\footnote{Present address: Dept. of
Physics and Astronomy, University College London, Gower St.,
London WC1E 6BT, England},
Gilles Montambaux$^1$ and Yuval Gefen$^2$}
\address{$^1$Laboratoire de Physique des Solides,
associ\'e au CNRS \\ Universit\'{e} Paris--Sud, 91405 Orsay, France}
\address{$^2$Department of Condensed Matter Physics,\\ The Weizmann Institute
  of Science, 76100 Rehovot, Israel}
\date{\today}
\maketitle
 \begin{center}
 \begin{abstract}
We investigate the addition spectrum of disordered quantum dots containing
spinless interacting fermions using the self-consistent Hartree-Fock
approximation. We concentrate on the regime $r_s\agt 1$, with finite 
dimensionless conductance $g$. We find that in this approximation
the peak spacing fluctuations
do not scale with the mean single particle level spacing for either Coulomb
or nearest neighbour interactions when $r_s\agt 1$. We also show that
Koopmans' approximation to the addition spectrum can lead to errors that
are of order the mean level spacing or larger, both in the mean addition
spectrum peak spacings, and in the peak spacing fluctuations.
\end{abstract}
\end{center}
\pacs{PACS Numbers: 73.20Dx, 73.23Hx}
%
\section{Introduction}
%
We consider the response of the ground state of spinless fermions to the
addition or removal of a particle. To this end, we apply the self-consistent
Hartree-Fock (SCHF) approximation: a non-perturbative effective single
particle theory.

Koopmans' theorem \cite{Koopmans} states that the single particle SCHF
energy levels describe the affinity and ionisation energy spectra for the
unoccupied and occupied states respectively.
The approximation involved is that all the other particles do not react
to this process. This approximation is generally considered to be good
when the single particle wavefunctions are extended: corrections to
each wavefunction due the rearrangement of the system following the
addition or removal of a particle are expected
to be ${\cal O}(1/N)$, where $N$ is the number of
particles in the system \cite{AGD}. Moreover, these corrections to the
wavefunctions are expected to disappear in the limit of vanishing
disorder (having e.g. periodic boundary conditions), where the single
particle wavefunctions are free waves, as well as in the limit of vanishing
interaction. It is then loosely assumed that in the thermodynamic limit,
Koopmans' theorem becomes exact even for disordered systems, and should be
sufficiently accurate for mesoscopic samples. It is evident however, that if
the physical quantities at hand require an energy resolution of order
the  mean single particle level spacing, $\Delta$, the validity of
Koopmans' theorem should be reconsidered.

Our analysis of Koopmans' theorem for a quantum dot is closely related
to the addition spectrum of the latter: the spectrum of energy
differences between
states with total particle number different by unity. We consider only
the energy differences between ground states, which is experimentally
accessible through resonant tunnelling \cite{Sivan96,Simmel,Marcus97},
and capacitance \cite{Ashoori92,Ashoori97} measurements at low bias and
temperature. We stress that while our analysis here pertains to some aspects of
the experiments, a direct comparison is not feasible: firstly we consider
spinless electrons, and secondly we consider disordered systems in the
diffusive regime; in Ref.\cite{Sivan96} the mean free path is of order
the sample size, whereas in
Ref.\cite{Marcus97} the mean free path is much larger than the system
size, and ergodicity is ensured by a chaotic boundary shape.

The position of the observed resonant tunnelling (RT) conductance peaks
can be related to the ground state energy difference
$\mu_N=E_G(N,V_G) - E_G(N-1,V_G)$ where $E_G(N,V_G)$ is the ground state
energy of the dot with $N$ electrons, at gate voltage $V_G$.
The spacing between consecutive peaks is thus related to
\begin{equation}
\Delta_2(N)= E_G(N+1,V_G)-E_G(N,V_G)-E_G(N,V'_G) + E_G(N-1,V'_G)\ .
\label{D2VG}
\end{equation}
Within the constant interaction (CI) model the ground state energy
is simply the sum of filled single particle energies $e(n)$ plus
$N(N-1)V_0/2$, where $V_0$ is the constant interaction. Taking
$V_G=V'_G$, the peak spacing trivially reduces to
\begin{equation}
\Delta_2(N)= e(N+1)-e(N)+V_0
\label{V0}
\end{equation}
 and so, in the diffusive regime,
displays shifted Wigner-Dyson (WD) statistics \cite{Efetov83,Mehta}
up to corrections in one over the dimensionless conductance, $g$
\cite{Kravstov}: $P(s)= \pi s / 2 \exp(-\pi s^2 /4)$ for zero
magnetic field, the case that we consider here;
$s=(\Delta_2 -V_0)/\Delta$.

Recent experiments on quantum dots \cite{Sivan96,Simmel,Marcus97}
have shown that whilst the mean peak spacings are well described by
the CI model, the fluctuations are not described by Wigner-Dyson statistics.
It is found that the distribution of $\Delta_2$ is roughly Gaussian 
\cite{Sivan96,Marcus97}, with broader non-Gaussian tails seen in
Ref.\cite{Marcus97}. In Refs.\cite{Sivan96,Simmel} the variance of the
fluctuations was found to be considerably larger than that given by the WD
distribution. Further experimental observations, including correlations
of peak heights \cite{Marcus96,Marcus98,Patel} and the sensitivity to
an Aharanov-Bohm flux \cite{Chang95,Marcus96,Berkovits98}
are not consistent with random matrix results, and suggest a breakdown
of the naive single particle picture. 
To investigate this, one needs information on the ground
state wavefunction.

Blanter {\it et al} \cite{Blanter97} have
evaluated the fluctuations within a Hartree-Fock framework, neglecting
effects due to the change in gate voltage $V_G$ to $V'_G$.
To this end they applied the Random Phase Approximation (RPA) to generate
the screened interaction in the confined geometry, and assumed
that all HF level spacings, except for the Coulomb gap, are described
by WD statistics. Implicitly assuming Koopmans'
theorem \cite{Koopmans} to be valid, and using wavefunction statistics
established for non-interacting electrons in a random potential,
they calculated the fluctuations of $\Delta_2$ beyond the CI model.
These additional fluctuations were found to be
parametrically small (in $1/g$) and proportional to $\Delta$.
Hence, the total fluctuations in $\Delta_2$ were found to be proportional
to $\Delta$. The analysis of Ref.\cite{Blanter97} is consistent in
the limits $g\gg 1$, $r_s\ll 1$. The parameter $r_s$ characterises the
relative importance of interactions in the electronic system, and
is defined as the mean electron separation in units of the effective
Bohr radius.

Exact numerical calculations on small disordered dots \cite{Sivan96}
did produce large Gaussian distributed fluctuations at experimental densities.
It was claimed that for strong enough interaction
$\delta \Delta_2 /\langle \Delta_2 \rangle$ is $\it universal$,
independent of the interaction strength and disorder, where
$\delta \Delta_2$ denotes the typical (RMS) size of the fluctuations, and the
angle brackets denote disorder averaging. This $\it universal$ constant
was found to be approximately $.10 - .17$, to be compared
with the WD result for the CI model: $.52 \Delta/(\Delta+V_0)$. We note that
the typical experimental value for the charging energy $V_0$, is much
larger than $\Delta$. The scaling with $\langle \Delta_2 \rangle$
suggested by this analysis \cite{Sivan96} is in stark contrast to the
scaling with $\Delta$ obtained in Ref.\cite{Blanter97}.

Stopa has considered ballistic chaotic billiards numerically, using
local density functional theory \cite{Stopa}. In this case, it was
claimed that the fluctuations arise due to strongly scarred wavefunctions
in the self consistent potential. As a result of these scars,
an asymmetric distribution of $\Delta_2(N)$ was found, including
strong correlations over $N$. It was then further noted that what is
actually measured (i.e. the change in the gate voltage between resonant
tunnelling peaks) is not simply related
to $\Delta_2(N)$ when the dependence of $\Delta_2(N)$ on the gate
voltage is strong. It was then claimed that
a self-consistent calculation of $\Delta V_G$ with $\Delta_2(N)$
retrieves a symmetric distribution of $\Delta_2(N)$,
and reduces peak to peak correlations.

A further suggestion that the coupling to the gate is important
in understanding the fluctuations has been made with reference to the
CI model, with WD statistics for the single particle levels
\cite{Vallejos98}. The authors claim that the
required distribution of $\Delta_2$ can be generated, except for
the non-Gaussian tails, through the de-correlation of neighbouring levels
under a parametric change in the Hamiltonian (mediated by $V_G$).
However, the degree of de-correlation induced by $\Delta V_G$ is left as
a fitting parameter.

In this paper we present numerical calculations within the SCHF
approximation, considering larger samples than is feasible by exact
diagonalisation \cite{Sivan96}. This approximation has been seen to
be quite good for the calculation of persistent currents in similar
systems \cite{Poilblanc}.
We show that fluctuations large compared to the single particle level spacing
can arise without recourse to varying the sample shape, size or gate to dot
coupling, supposing these to be {\it additional} effects. We further
demonstrate that approximating the addition spectrum spacings by applying
Koopmans' theorem can lead to large errors in the calculation of the
spacing statistics.

We consider separately both a long range (Coulomb) bare interaction and
a short range (nearest neighbour) bare interaction.
In section \ref{Model} we introduce our model in detail;
in section \ref{Implications} we present a short discussion of the
implications of Koopmans' theorem; in section \ref{Results} we present
and discuss our numerical results,
which are then summarised in the final section.
%
\section{The Model}
\label{Model}
%
We address the following tight binding Hamiltonian for spinless
fermions
\begin{equation}H = \sum_i w_i c^+_i c_i - 
t\sum_{i, \eta}c^+_{i+\eta} c_{i} +
{U_0\over 2} \sum_{ij} M_{ijij} c^+_i c^+_j c_j c_i
\label{H}
\end{equation}
where $i$ is the site index,
$\eta$ describes the set of nearest neighbours, $w_i$ is the
random on site energy in the range $[-W/2,W/2]$, and
$t$ the hopping matrix element, henceforth taken as unity. We study
separately, both a Coulomb interaction potential,
\begin{equation}
M_{ijij}= 1/|{\bf r}_i-{\bf r}_j|
\end{equation}
and a short range potential plus a constant term $M_c$ (see below),
\begin{equation}
M_{ijij}=(\delta_{i,i+\eta}+M_c)\ .
\label{srmel}
\end{equation}
We consider a 2D system with periodic boundary conditions, and choose to
define 
\begin{equation}
|{\bf r}_i-{\bf r}_j|^2 \equiv [L_x^2\sin^2 (\pi n_x /L_x)+
L_y^2\sin^2(\pi n_y/L_y)]/\pi^2\ ,
\end{equation}
where $(n_x,n_y) \equiv {\bf r}_i-{\bf r}_j$.

All distances are measured in units of the lattice spacing $a$; the
physical parameters are therefore $U_0=e^2/a$, $t=\hbar^2/2ma^2$.
The standard definition for $r_s$ is given,
for low filling, by $r_s=U_0/(t\sqrt{4\pi\nu})$, where $\nu=N/A$ is the
filling factor on the tight binding lattice with $A$ sites.
The dimensionless conductance $g$ can be approximated, again for low
filling, using the Born approximation. We find
$g= 96\pi\nu(t/W)^2$, which is valid for $A,N\gg g\gg 1$.
Here $\nu\approx 1/4$ throughout, so that $r_s\approx .56 U_0/t$, and
$g\approx 75(t/W)^2$.

Having identified the parameters of our model with the standard ones
employed in the theory of a continuous electron gas, we note that 
in the limit of small $r_s$ and $1/g$
the leading order term for the typical interaction dependent fluctuations
predicted by Blanter {\it et al} \cite{Blanter97} is 
$\sim U_0\Delta/t\sqrt{g}$. For the torus geometry considered here, this
contribution, being a surface term, vanishes identically. Their
prediction then reduces to typical fluctuations in addition to
those of the CI model to be of order $U_0 \Delta/tg$.

The torus geometry has the advantage over geometries with hard walls whereby
in the former, the compensating background charge provides a trivial
shift in all the site energies, and can be removed. In a bounded dot,
with an overall charge, the excess charge may build up near the
boundary, depending on the position of nearby metallic plates and gates.
These effects are geometry specific \cite{Blanter97}.
Upon adding an electron, the average charge configuration may change
considerably (the configuration is strongly geometry dependent).
As the gate voltage is varied to allow for the next electron addition, the
background potential could have changed causing further charge rearrangement.
Whilst it is of
great interest to analyse this issue (which may play an important role in
the peak spacing fluctuations as well as undermining the naive
single particle picture by further reducing the accuracy of
Koopmans' theorem \cite{Patel}), we concentrate here on 
effects due entirely to the {\it intrinsic rearrangement} of the dot.
From this point of view our analysis may be taken as an attempt to
establish an upper bound criterion for the breakdown of Koopmans' theorem.
In reality it may break down earlier due to other non-universal factors. During
the completion of this work very recent experimental evidence for significant
rearrangement has been produced \cite{Patel}. It is argued that rearrangements
due to adding an electron are far greater than rearrangements due merely to
a change in shape.

When considering the short range interaction, the mean charging energy
$V_0$ in Eq.(\ref{V0}) must be put in by hand through $M_c$ of
equation (\ref{srmel}). The way in which this is done depends on the
physical situation being modelled, and is highly geometry dependent
(vis-\` a-vis the gates). We stress that the value of $M_c$ does not
affect the physical results. We choose to insert
\begin{equation}
M_c=V_0/U_0-4/A\ ,\qquad
V_0={\sum_{{\bf r}\ ,{\bf r'}}}' {U_0 \over |{\bf r} -{\bf r'}|}.
\label{Mc}
\end{equation}
This value for $M_c$ is defined such that if the charge is uniformly spread
over the dot, the average charging energies in the Coulomb and
nearest-neighbour cases roughly coincide. This choice has been made
for simplicity, but corresponds
to the premise that the interactions of the $N$-electron gas with the positive
background and with itself is the same for both models considered. Exchange
contributions, which tend to reduce the total charging energy, are included
insofar as to cancel both the on-site contributions to the energy, and the
unphysical self-interaction of electrons, but are otherwise
neglected \cite{Blanter97,BerkAlt}. The energy associated with charging the
system uniformly is
$U_0 N(N-1)/(2A^2)\sum_{{\bf r}\ ,{\bf r'}}' |{\bf r} -{\bf r'}|^{-1}$
in the Coulomb case, and
$U_0 N(N-1)/(2A^2) (4A +M_c A^2)$ in the nearest-neighbour case.
Equation (\ref{Mc}) follows from equating these energies.
This estimate can be systematically improved if the above premise is taken
as the definition, not only by correctly accounting for the exchange
contributions, but also by considering single particle wavefunction statistics
in the diffusive regime. In this case wavefunction correlation functions
such as $\langle |\psi_i ({\bf r})|^2  |\psi_j ({\bf r'})|^2 \rangle$ are
required in order the calculate the average electrostatic energy, where
here and after $\langle\dots\rangle$ denotes averaging over the disorder
ensemble.

In Ref.\cite{Blanter97} it was assumed that the (RPA) screening can be taken
into account before constructing the Slater determinant ground state, and
therefore their result corresponds to a short-ranged effective interaction.
It is not clear that this
remains a consistent procedure when calculating the ground state energy
self-consistently. The reason for the inconsistency is that many of the
diagrams generated by the SCHF approximation are already included in the RPA
calculation of the screening, resulting in double counting.
On the other hand, if the screening is generated externally (e.g. by close
metallic gates), then it is consistent to insert a short ranged bare
interaction, and this is the point of view taken here.

In some sense, the Coulomb interaction results can be considered as the
opposite limit of the nearest-neighbour interaction, and is of interest
in this context. However, it is more difficult to physically motivate the
use of a Coulombic bare interaction unless one is considering very low
electron densities. Screening is indeed weak in a $2d$ electron gas in a
vacuum, even at high density, but the
SCHF procedure cannot correctly generate screening by itself; whilst
it can screen the Hartree contributions (as discussed above), it does
not screen the exchange (Fock) term. However, we have verified that for
the range of parameters considered here, fluctuations in the Hartree
energy are larger than the typical fluctuations of the exchange energy. This
suggests that the error made in not screening the exchange term correctly is
not overly important.
%
\section{Implications of Koopmans' Theorem}
\label{Implications}
\begin{figure}[h]
\centerline{
\epsfysize 5cm
\epsffile{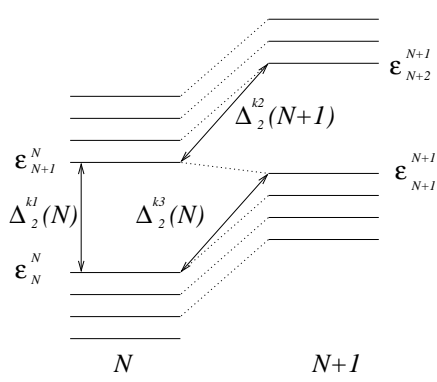}}
\caption{A schematic digram of the SCHF spectra of $N$ and $N+1$ particles.}
\label{spectra}
\end{figure}

%
Let us now consider the form of $\Delta_2$, and approximations to it given
by applying Koopmans' theorem. We denote the diagonal matrix elements of
the one-body operators in (\ref{H}) by $T_i^N$, and the
antisymmetrised Hartree-Fock interaction by $V_{ij}^N$, where hereafter
the subscripts denote single particle states, and the superscript $N$ denotes
the number of particles present and identifies the self-consistent basis
of single particle wavefunctions being employed, $\psi_i^N$.
For the torus geometry, where the gate voltage and background potential
represent a trivial shift that can be omitted, the SCHF ground state energy
is given by
\begin{equation}
E_G (N)=\sum_j^N \epsilon_j^N -{1\over 2}\sum_{ij}^N V_{ij}^{N}
=\sum_j^N T_j^N+{1\over 2}\sum_{ij}^N V_{ij}^{N}\ ,
\label{SCHFgsE}
\end{equation}
where $\epsilon_l^m$ is the $l$th SCHF single particle energy for a
system of $m$ particles in the ground state:
\begin{equation}
\epsilon_l^m=T_l^m + \sum_j^m V_{l\,j}^m \ .
\label{eHF}
\end{equation}
Using (\ref{SCHFgsE}), we find (c.f. Eq.(\ref{D2VG}))
$$
\Delta_2(N)=T_{N+1}^{N+1} - T_{N}^{N-1} +
\sum_j^N \Big ( T_{j}^{N+1} - 2  T_{j}^{N} + T_{j}^{N-1} \Big )
\qquad\qquad
$$
\begin{equation}
+ \sum_j^N \Big ( V_{N+1\,j}^{N+1} - V_{N\,j}^{N-1} \Big ) +
{1\over 2}\sum_{ij}^N\Big ( V_{ij}^{N+1}-2V_{ij}^{N}+V_{ij}^{N-1}\Big )
\ .
\label{delta2}
\end{equation}
Applying Koopmans' approximation corresponds to dropping the
superscripts and employing an appropriate fixed basis.
The theorem implies that the effective single particle {\it states}
do not depend on the occupation of these states.
In particular, Koopmans' theorem yields
$\epsilon_{N+1}^N$ for the minimum energy required to add a particle
to a system of $N$ particles, and $\epsilon_N^N$ for the maximum energy gained
by removing a particle from the same system; in both cases the final state
is a ground state.
Clearly $\epsilon_l^m$ as well as the ground state energy depend on $m$,
even in Koopmans' approximation, through the number of terms in the sum
in Eqs.(\ref{eHF}) and (\ref{SCHFgsE}) respectively.
It is then easy to see that Koopmans' approximation yields
\begin{equation}
\Delta_2^{k_1}(N)=\epsilon^N_{N+1}-\epsilon^{N}_{N}\ .
\label{k1}
\end{equation}
We also consider two other approximations to $\Delta_2$ that
involve calculating two self-consistent bases rather than just one:
\begin{equation}
\Delta_2^{k_2}(N)=\epsilon^N_{N+1}-\epsilon^{N-1}_{N}
\label{k2}
\end{equation}
\begin{equation}
\Delta_2^{k_3}(N)=\epsilon^{N+1}_{N+1}-\epsilon^{N}_{N}\ .
\label{k3}
\end{equation}
All three estimates (\ref{k1}-\ref{k3}) coincide with
$\Delta_2(N)$ of Eq.(\ref{delta2}) if Koopmans' theorem holds.
To connect with the notation of Ref.\cite{Blanter97}, and to
demonstrate the difference between the above three approximations
and the fully self-consistent result, we provide a schematic diagram
of the SCHF spectra in Fig. \ref{spectra}.

Since the self-consistent basis of $N$ particles provides the
lowest energy for $N$ occupied levels, and similarly for $N-1$ particles,
the following relations are clear:
$$
\sum_i^{N-1} T_i^N + {1\over 2}\sum_{ij}^{N-1} V_{i\,j}^N \ge
\sum_i^{N-1} T_i^{N-1} + {1\over 2}\sum_{ij}^{N-1} V_{i\,j}^{N-1}
$$
\begin{equation}
\sum_i^{N} T_i^N + {1\over 2}\sum_{ij}^{N} V_{i\,j}^N \le
\sum_i^{N} T_i^{N-1} + {1\over 2}\sum_{ij}^{N} V_{i\,j}^{N-1}
\label{inequal}
\end{equation}
Combining these equations, we find that
$\Delta\epsilon(N)\equiv\Delta_2^{k_1}(N)-\Delta_2^{k_2}(N)\ge 0$,
or equivalently
\begin{equation}
\Delta\epsilon(N)\equiv\epsilon_N^{N-1}-\epsilon_N^N \ge 0\ .
\label{deltae}
\end{equation}
The equalities in (\ref{inequal}),(\ref{deltae}) only hold when no
modification of the effective single
particle wavefunctions occurs following the addition of an electron. In a
disordered dot, in which there are no spatial symmetries, such a modification
will always take place, and so $\Delta\epsilon$ can be considered strictly
positive.

\begin{figure}[h]
\centerline{
\epsfysize 5cm
\epsffile{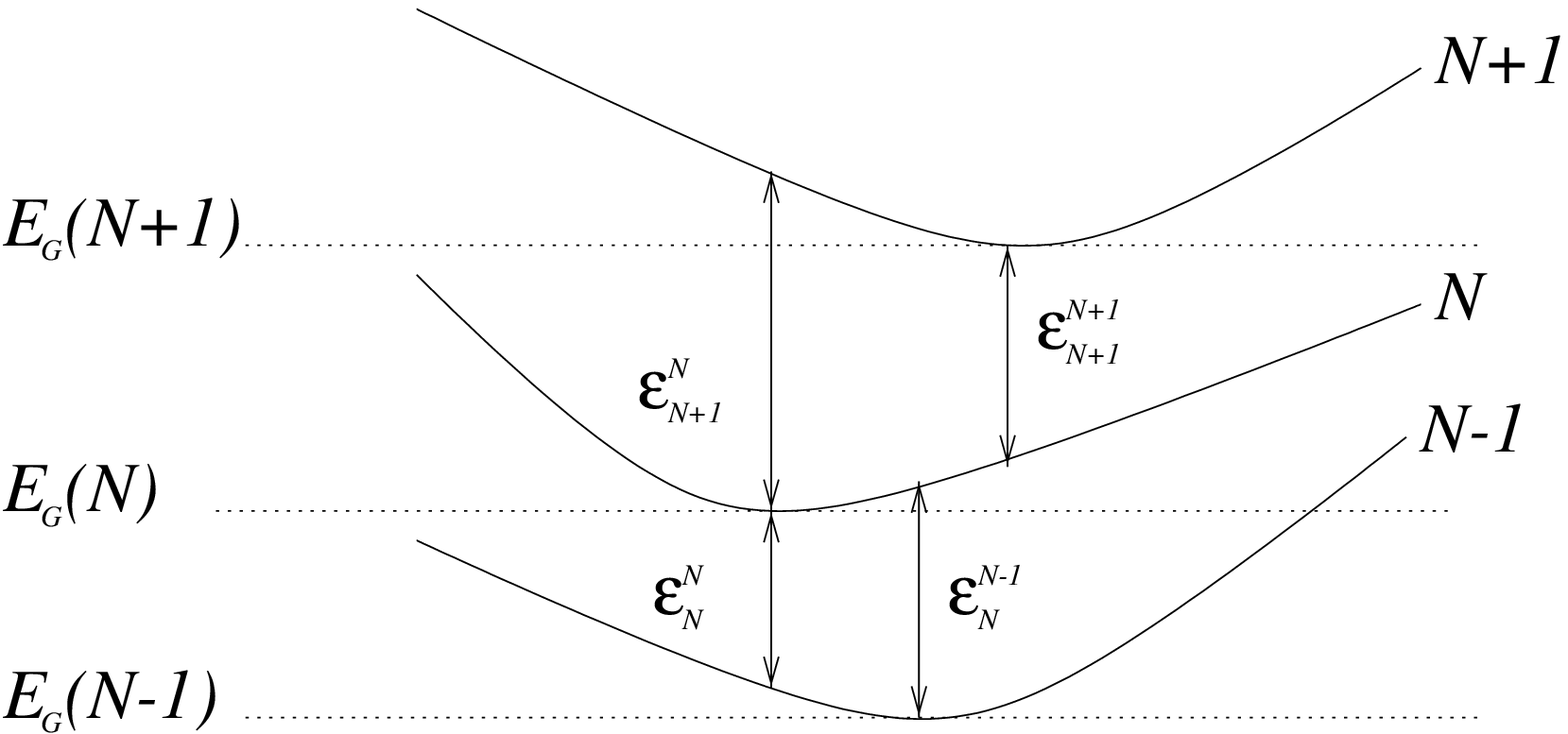}}
\caption{A schematic diagram representing the expectation value of the 
Hamiltonian in the spaces of $N-1$, $N$ and $N+1$ Slater determinants
(superimposed). Distances are only meaningful within a given space.
In order to define distances between two Slater determinants in two
different spaces, i.e. $\Psi^N$ and $\Psi^{N+1}$, we include the first
unoccupied state with $\Psi^N$ \protect\cite{Distance}. In this way Koopmans'
theorem is exact when the two Slater determinants coincide.
The SCHF solutions correspond to minima
in these surfaces. It can be seen that the Koopmans' approximations
$\epsilon^N_{N+1}$ and $\epsilon^N_N$ to the addition energy
$E_G (N+1) -E_G (N)$ are upper and lower bounds respectively. Similarly
for the addition energy $E_G (N) -E_G (N-1)$. The definition of the
Koopmans' approximation to $\Delta_2$ given by Eq.(\protect\ref{k1}) can be
seen to contain the difference between these bounds.}
\label{quadE}
\end{figure}

The difference $\Delta\epsilon$ provides a measure of the effectiveness of
Koopmans' theorem. To demonstrate this we present, in Fig.\ref{quadE},
a schematic diagram of the
surface of expectation values of the many-body Hamiltonian in the space
of Slater determinants of $N-1$, $N$, and $N+1$ particles. The SCHF ground
states correspond to minima in these surfaces. From the diagram, it is
clear that the energies $\epsilon^N_{N+1}$ and $\epsilon^{N-1}_N$
are upper bounds to the respective addition energies, and the
energies $\epsilon^{N+1}_{N+1}$ and $\epsilon^N_N$ are lower
bounds to the addition energies. The approximation $\Delta_2^{k_1}$,
is therefore obtained by subtracting a lower bound ($\epsilon^N_N$)
from an upper bound ($\epsilon^N_{N+1}$). As a result, the average value
contains the average difference between
the two bounds in addition to the correct mean $\Delta_2$.
It is generally assumed that the difference between
the two bounds vanishes in the thermodynamic limit, and therefore
so does $\Delta\epsilon$. We shall see that our results do not show
any indication that this
is the case. On the other hand, $\Delta_2^{k_2}$ corresponds to the difference
of two upper bounds to the two relevant addition energies. Regardless of the
quality of the upper bound, so long as it is not strongly dependent on the
number of particles present, both the particle number and disorder averaged
results are good. The third approximation to $\Delta_2^{k_3}$ corresponds to
the difference of two lower bounds, and like $\Delta_2^{k_2}$ is good in
the mean. It is for this reason that we introduce these alternative
approximations. It is 
easy to see that $\Delta_2^{k_1}(N)-\Delta_2^{k_3}(N)=\Delta\epsilon (N+1)$
and therefore provides no further information. On the other hand, the
fluctuations of $\Delta_2^{k_3}$ can be different from those of
$\Delta_2^{k_2}$, and so are investigated separately. We note that
in a clean system at $r_s$ below the Wigner crystal transition
\cite{Ceperley}, the minima would align in Fig. \ref{quadE}, reflecting the
validity of Koopmans' theorem in that limit.

Let us briefly discuss the non-self-consistent
single particle picture, for which the Koopmans' approximations
(\ref{k1}-\ref{k3}) and Eq.(\ref{delta2}) all coincide: 
\begin{equation}
\Delta_2 (N)= T_{N+1}-T_N+\sum_j^{N-1} \Big ( V_{N+1\,j} - V_{N\,j} \Big )
+V_{N+1\,N}\ .
\label{noSC}
\end{equation}
Here, the term {\it non-self-consistent approximation} refers to a scheme
where a set of effective single-particle states is given (e.g. by solving
the N-electron SCHF problem), and utilised for any number of particles
present in the system.
The nearest neighbour spacings between levels that are both occupied or
unoccupied has a similar form:
\begin{equation}
\epsilon_{m+1}^N-\epsilon_m^N=
T_{m+1}-T_m+\sum_j^{N} \Big ( V_{m+1\,j} - V_{m\,j} \Big )\ ,
\label{nnspace}
\end{equation}
the major difference between (\ref{noSC}) and (\ref{nnspace}) is the additional
{\it unbalanced} matrix element $V_{N+1\,N}$ appearing in (\ref{noSC}).
Let us also suppose that in this simple single particle scheme the electrons
interact with a short-ranged effective interaction.
Blanter {\it et al} \cite{Blanter97} introduce the hypothesis
that the (normalised) spacings (\ref{nnspace}) and $\Delta_2-V_{N+1\,N}$
obey WD statistics up to corrections in $1/g$.
Further assuming that the wavefunction
correlations are still close to those of non-interacting particles
leads, for the short-ranged effective interaction,
to the result ${\rm Var}(V_{ij})\sim (U_0\Delta/tg)^2$ \cite{Blanter},
so that the interaction dependent contribution to $\delta\Delta_2$
scales like $U_0\Delta/tg$ \cite{Blanter97}. This analysis is valid
in the regime $r_s \ll 1$ and  $g\gg 1$, implying that
Koopmans' theorem is a good approximation in that regime.
%
\section{Results and Discussion}
\label{Results}
%
In this section we present and discuss the results of the numerical simulations
for both the nearest-neighbour and the Coulomb bare potentials. To make each
subsection self-contained there is some repetition.
%
\subsection{Short Range Interactions}
\begin{figure}[h]
\centerline{(a)
\epsfysize 5cm
\epsffile{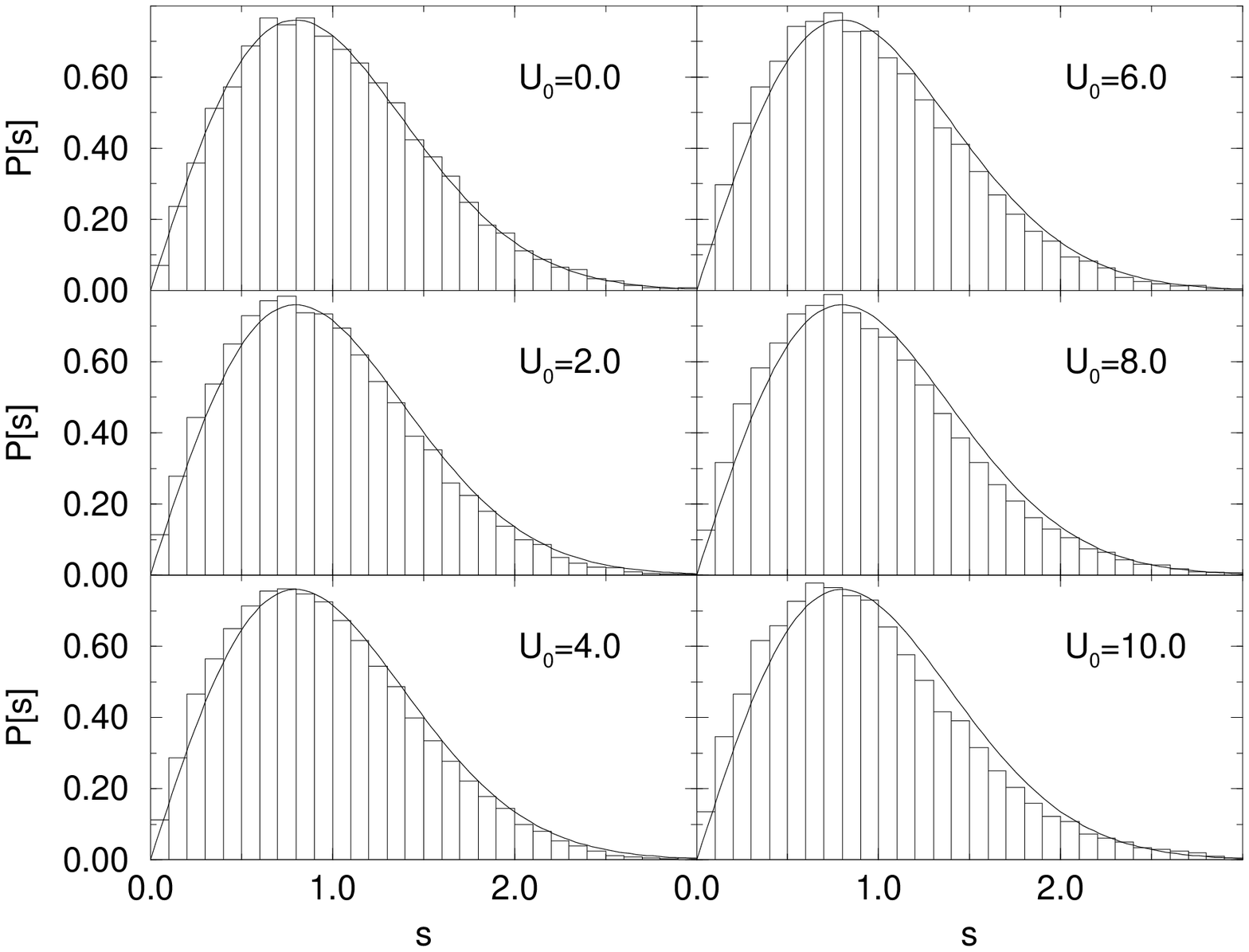}
\hspace{.5cm}(b)
\epsfysize 5cm
\epsffile{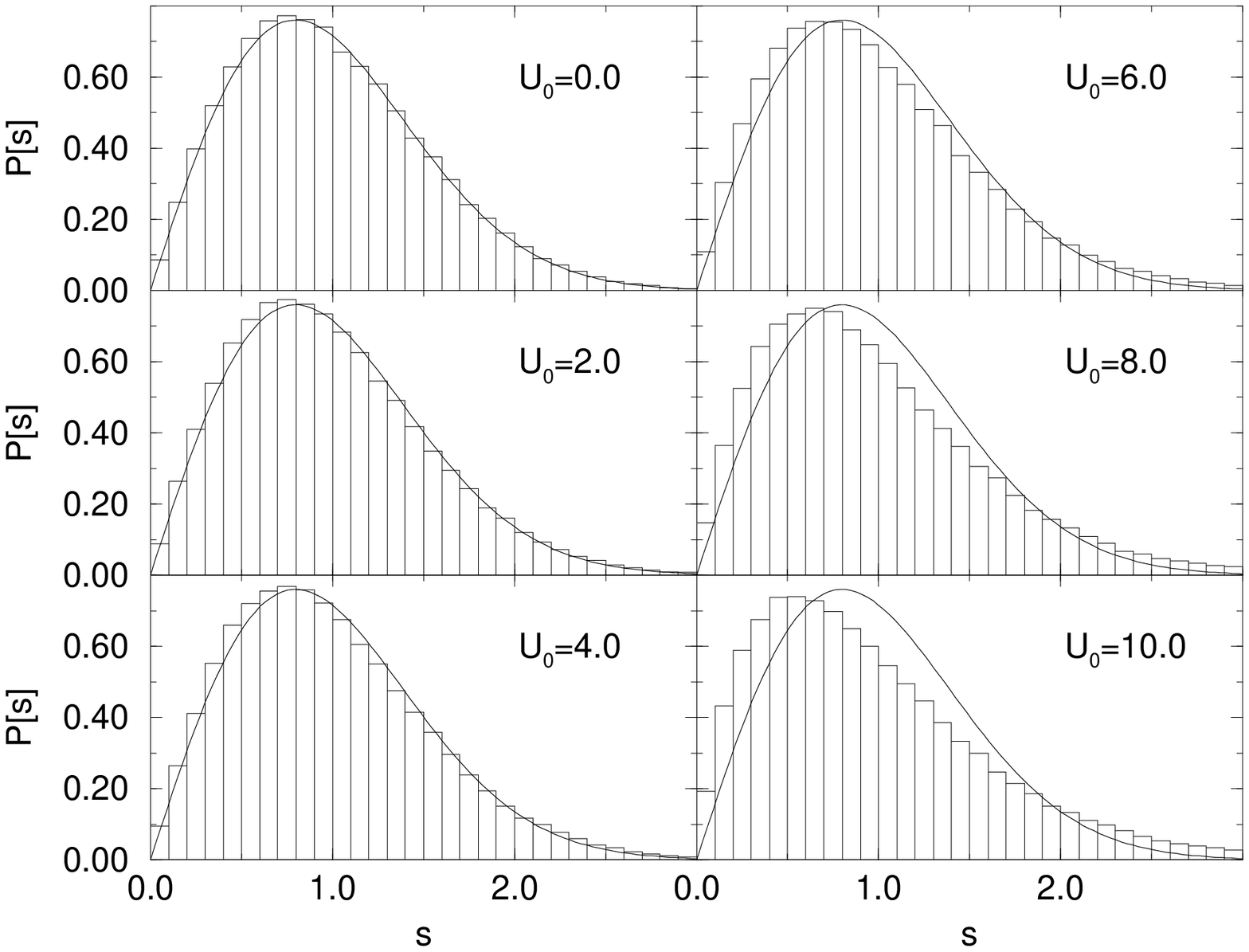}
}
\centerline{(c)
\epsfysize 5cm
\epsffile{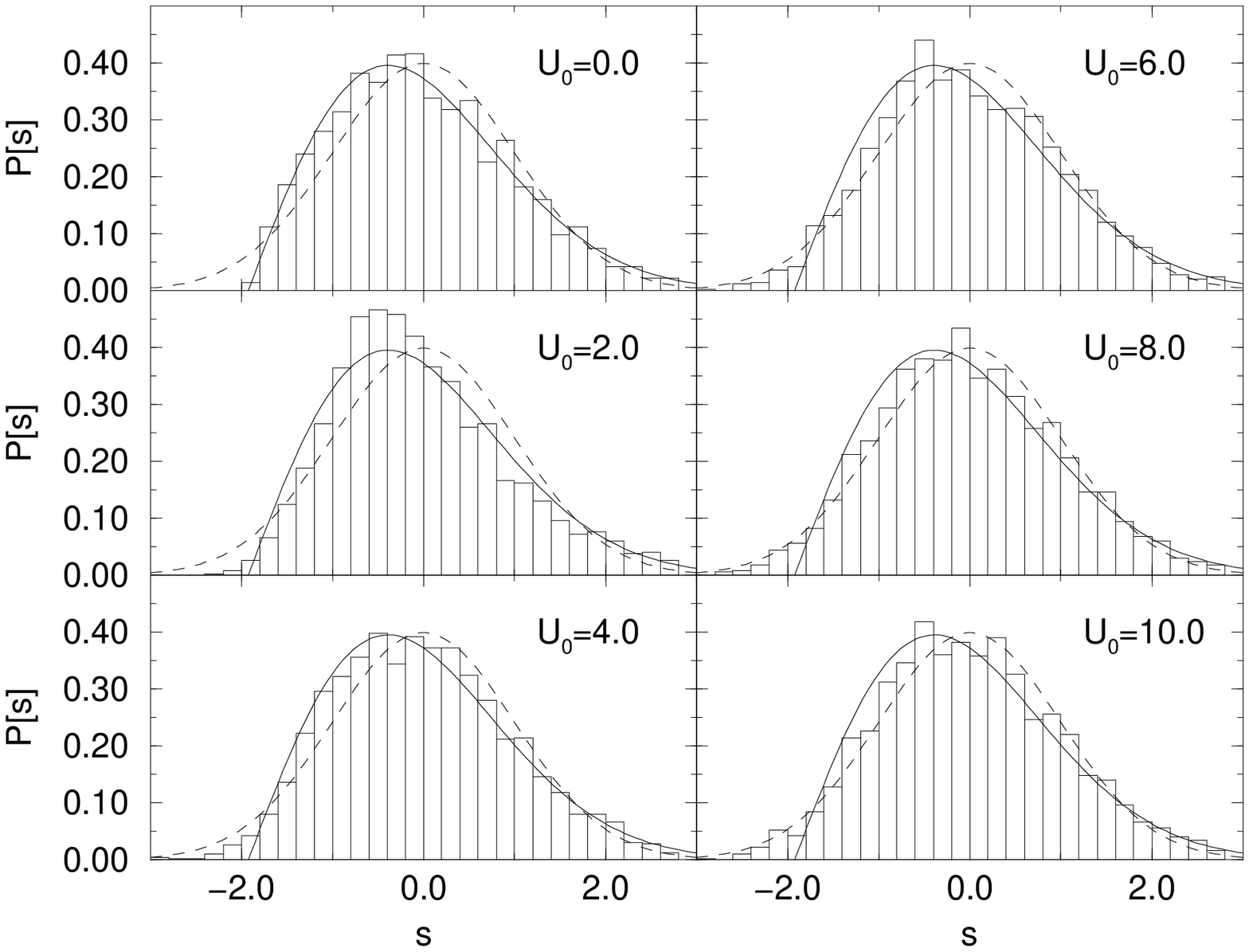}
}
\caption{SCHF level spacing distributions $P[s]$ for a) occupied states,
b) unoccupied states; $s=\Delta E/\langle\Delta E\rangle$, and c) the gap
$\Delta_2^{k_1}$, where
$s=(\Delta_2^{k_1} - \langle\Delta_2^{k_1}\rangle)/\delta\Delta_2^{k_1}$.
The solid lines show the WD distribution,
and in (c) the dashed line follows a Gaussian law. The samples were $8*9$
lattices with 14 electrons and nearest-neighbour interactions; $W=2$.
$r_s \approx 0.56 U_0/t$.
The statistics were obtained from an ensemble of 2500 samples.}
\label{pssr}
\end{figure}

%
We consider first the case of a nearest neighbour bare interaction potential
as defined in Eq.(\ref{srmel}).
We begin by plotting the distributions of both the level spacings 
(\ref{nnspace}) and the gap (\ref{noSC}) of the SCHF spectrum for finite $r_s$,
$g$. In this case (\ref{noSC}) and (\ref{nnspace}) are calculated
in the self-consistent basis of $N$ particles.  In Fig. \ref{pssr}
it is seen that
the normalised level spacings between occupied states show an increasing
deviation from WD to Poisson statistics as $U_0$ is increased. This is
also true for the unoccupied states, but to a much greater extent. The 
difference between occupied and unoccupied states in the SCHF approximation
will be discussed in greater detail later. We interpret the tendency towards
Poisson statistics as a signature of the incipient
localisation of the effective one particle states.

The normalised gap ($\Delta_2^{k_1}$) distribution tends towards a
more symmetric distribution that is approximately
Gaussian as $U_0$ is increased.

\begin{figure}[h]
\centerline{
\epsfysize 7.5cm
\epsffile{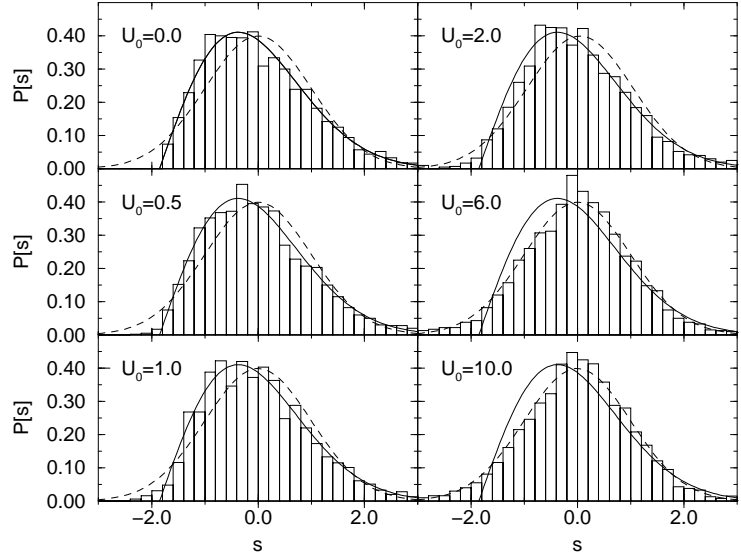}}
\caption{Distributions $P[s]$ where
$s=(\Delta_2-\langle\Delta_2\rangle)/\delta\Delta_2$
for various interaction strengths obtained self consistently for
the nearest neighbour interaction. The solid line show the WD distribution,
the dashed line shows the Gaussian distribution. The samples were
7*8 lattices with $N=15$, $W=4$, and the  statistics were
obtained from an ensemble of 3000 samples. $r_s \approx 0.56 U_0/t$.}
\label{pd2sr}
\end{figure}

We have also investigated the gap ($\Delta_2$) distribution obtained within
the fully self-consistent scheme, which we show in Fig. \ref{pd2sr}. We find
that as $U_0$ is increased, the distribution evolves from a WD form to
a more symmetric distribution similar to a Gaussian.

\begin{figure}[h]
\centerline{
\epsfysize 7.5cm
\epsffile{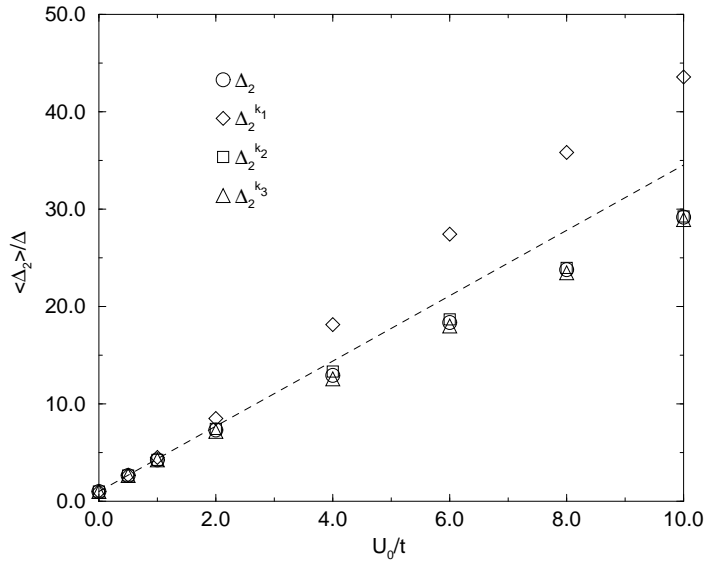}}
\caption{Typical result for the mean Coulomb gap averaged over 400
disorder realisations. Here, $W=4$, the lattice is $9*8$ and $N=15$. The dashed
line is the CI result. $r_s \approx 0.62 U_0/t$.}
\label{srexm}
\end{figure}

We shall concentrate first on the mean values of these distributions.
A typical dependence of $\langle \Delta_2\rangle$ on the interaction
parameter $U_0$ is plotted in figure \ref{srexm}.
Whilst $\langle\Delta_2^{k_2}\rangle$ and
$ \langle\Delta_2^{k_3}\rangle$ provide a good approximation,
we see a strong deviation of $\langle\Delta_2^{k_1}\rangle$.
For all the system sizes considered, this effect occurs at
$r_s\sim {\cal O}(1)$.
Results for the CI model, evaluated as described above, are
plotted for comparison. Deviations of order ${\cal O}(U_0/A)$ from the
CI model appear above $U_0\approx 2$ ($r_s\approx 1$).

Elsewhere \cite{Walker}, we show that the ground state develops large
density modulations as $U_0$ is increased beyond $r_s \sim{\cal O}(1).$
These ground state charge density modulations (CDMs) explain the
deviations of $\langle\Delta_2\rangle$ from the CI model prediction:
they reduce the average addition energy by up to $4U_0/At$ when
$\nu< 1/2$ for a commensurate lattice. When $\nu>1/2$  the mean charging
energy can be correspondingly increased.

We are also now in a position to understand why the level spacing
statistics between unoccupied states show an increased tendency towards
a Poisson distribution: the density modulations that appear in the
ground state alter the potential felt by the unoccupied states.
These modulations are not spatially ordered, as is demonstrated in
Ref.\cite{Walker}. Hence, as $U_0$ is increased, the unoccupied states see
an effective potential with increasingly strong modulations and tend to
localise, whence the tendency towards a Poissonian distribution.

\begin{figure}[h]
\centerline{
\epsfysize 7.5cm
\epsffile{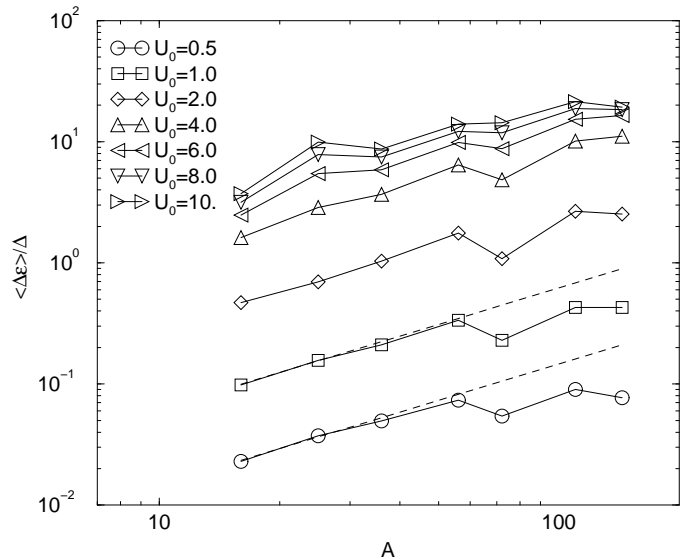}}
\caption{$\langle\Delta\epsilon\rangle/\Delta$ against the sample area
$A$ after 75 to 1000 disorder realisations. The dotted lines are
proportional to $A$ as guides for the eye. $r_s \approx U_0/2t$.}
\label{desra}
\end{figure}

Let us consider the error in $\langle\Delta_2^{k_1}\rangle$ in more detail.
As can be seen in Fig. \ref{desra}, $\langle\Delta\epsilon\rangle/\Delta$
increases with the system size for lattices up to about 7*8; for larger
systems ($A\agt 50$)  it seems that the error becomes proportional to
$\Delta$, and is of order $\Delta$ when $r_s$ is of order unity.

\begin{figure}[h]
\centerline{
\epsfysize 7.5cm
\epsffile{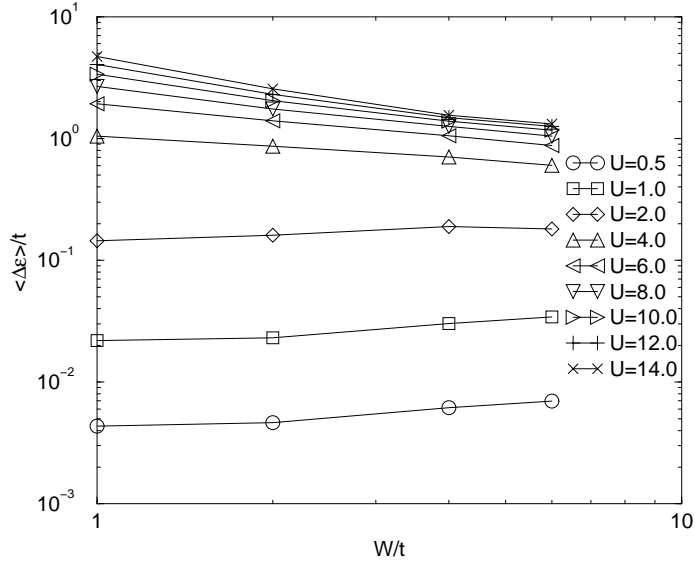}}
\caption{$\langle\Delta\epsilon\rangle/t$ against disorder $W$
averaged over 200 disorder realisations for a $11*10$ lattice with $N=28$.
$r_s \approx 0.54 U_0/t$.}
\label{desrw}
\end{figure}

We find that the nature of the disorder dependence of
$\langle\Delta\epsilon\rangle$ depends on the interaction strength,
as seen in Fig.
\ref{desrw}. The change in dependence occurs at interactions strengths 
corresponding to $r_s\approx 1$ for all the sample sizes considered.
One might be surprised that deviations from Koopmans'
theorem do not smoothly decrease as $W\to 0$ since,
in the limit of vanishing disorder, Koopmans' theorem becomes
exact on the torus due to the restoration of translational symmetry.
However, when $W\to 0$, the spectrum develops many near degeneracies such
that the effective perturbation due to $U_0$ is magnified as $W\to 0$.
In the limit $r_s\to 0$, $g\gg 1$, the typical size of the matrix
elements $V_{ijkl}$ which
drive the rearrangement scale like $\delta V_{ijkl}\sim r_s \Delta/g$
\cite{Blanter}, thus one expects that in this regime
$\langle\Delta\epsilon\rangle$ should increase with disorder.
We find only a weak increase with disorder for $r_s \alt 1$.

\begin{figure}[h]
\centerline{
\epsfysize 7.5cm
\epsffile{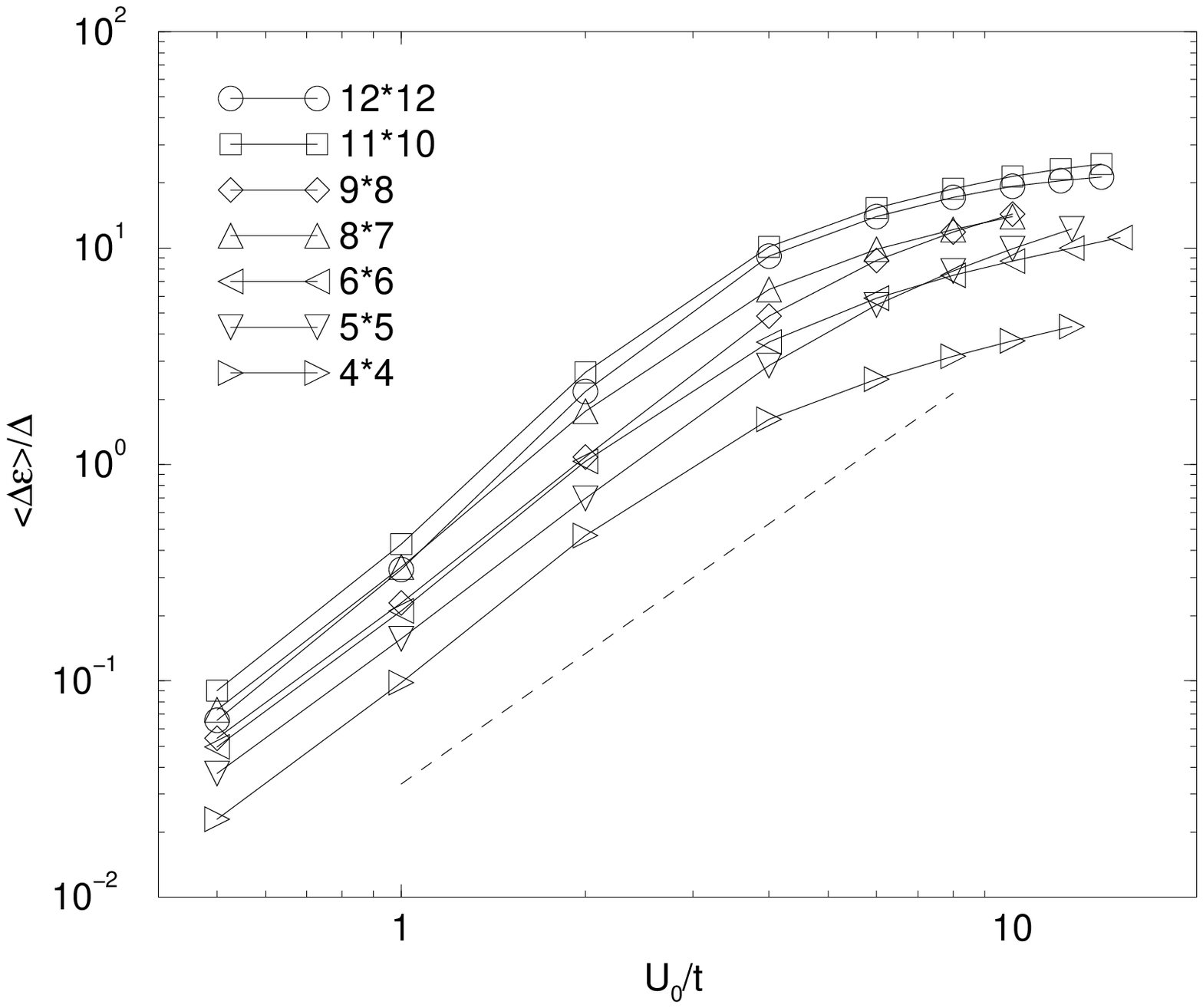}}
\caption{$\langle\Delta\epsilon\rangle/\Delta$ against $U_0$ after 75 to
1000 disorder realisations, for a range of sample sizes: $\nu\approx 1/4$,
on a log-log scale. The dashed line is a plot of
$\langle\Delta\epsilon\rangle/\Delta\propto U_0^2/t^2$, $r_s \approx U_0/2t$.}
\label{desru}
\end{figure}

In figure \ref{desru} we plot the interaction dependence of
$\langle\Delta\epsilon\rangle/\Delta$. We find that at small $U_0$,
$\langle\Delta\epsilon\rangle/\Delta\propto (U_0/t)^2$, with deviations
for larger $U_0$. In fact, this quadratic behaviour can be understood
using second order perturbation theory. To see this we refer back
to the schematic diagram of Fig. \ref{quadE}.
A shift \cite{Distance}
occurs in the ground state configuration when a particle is added, which is
represented by a misalignment of the minima. This shift is, to leading
order, linear in $U_0$. Since the SCHF ground state energy is a minimum in the
expectation value of the Hamiltonian, the difference of ground state
energies for the two configurations will be quadratic in this shift.
Furthermore, the local curvature tensor is independent of $U_0$ when the
interaction matrix elements are small compared to the mean level
spacing\cite{Thouless}. Thus, $\langle\Delta\epsilon\rangle$ scales
like $U_0^2$ in the perturbative regime. The indication is that second order
perturbation theory is qualitatively good even for $r_s \sim 1$.
We note that since both the
shift and the local curvature tensor depend on the disorder, there is
no such simple $W$ dependence.

To summarise the results for the mean Coulomb gap,
we find that Koopmans' approximation (\ref{k1}) makes an
error in the mean charging energy which for small $r_s$, $A\alt 50$,
and fixed disorder scales like $\langle\Delta\epsilon\rangle\propto r_s^2$.
There is also evidence that for the larger sizes ($A\agt 50$), far
beyond that accessible by exact diagonalisation, that
$\langle\Delta\epsilon\rangle\propto r_s^2 \Delta$.
The latter dependence is consistent with
the expectation that for sufficiently small $r_s$, $1/g$, perturbation
theory is valid when the effective interaction is short-ranged. We find
that $\langle\Delta\epsilon\rangle\sim{\cal O}(\Delta)$ when
$r_s \sim{\cal O}(1)$. To understand
this result, we return to (\ref{noSC}), (\ref{nnspace}), and fix the basis
to be the self-consistent one for $N$ particles, so that (\ref{noSC})
now describes $\Delta_2^{k1}$.
Since we have verified that the level spacings (\ref{nnspace}) show nearest
neighbour separation statistics which are close to WD for all $m\ne N$, 
with an approximately constant density of states, we are
led to conclude that $\langle\Delta\epsilon\rangle$ arises due to
the fundamental difference between occupied and unoccupied levels in the
SCHF approximation. In short, whilst
$\langle V^N_{m+1\, j} -V^N_{m\, j}\rangle$ for $m\ne N$ vanishes as
expected, $\langle V^N_{N+1\, j} -V^N_{N\, j}\rangle$ does not.
Indeed $\langle\Delta\epsilon\rangle\propto\sum_j^{N-1}
\langle V^N_{N+1\, j} -V^N_{N\, j}\rangle$.

\begin{figure}[h]
\centerline{
\epsfysize 7.5cm
\epsffile{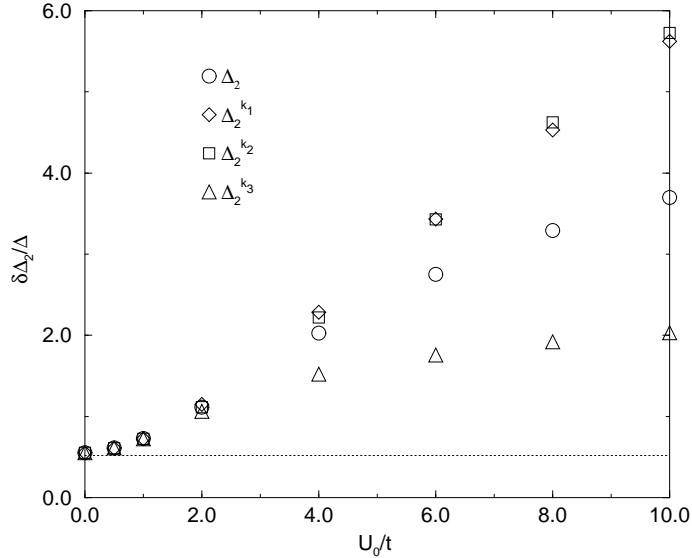}}
\caption{Typical result for the typical fluctuations of the Coulomb gap
averaged over 400 disorder realisations. Here, $W=4$, the lattice is $9*8$
and $N=15$.$r_s \approx 0.62 U_0/t$.}
\label{exsrv}
\end{figure}

Turning now to the fluctuations in $\Delta_2$, we plot an example
result in Fig. \ref{exsrv}. The most striking behaviour is the asymptotic
saturation of the fluctuations (verified but not shown for even stronger
interactions).
As with the deviations of $\langle\Delta_2\rangle$ from the CI model, this
occurs over the same range of interactions for all the system
sizes considered, and is associated with the appearance of CDMs.
Over the range of interaction strengths shown
the fluctuations have not completely saturated, but the ground state
density modulations are already present \cite{Walker}, such that,
at low filling, the short-ranged contribution to the interaction
energy is reduced. In the limit of strong interactions the charge
segregates at a kinetic energy cost of order ${\cal O}(t)$, and $U_0$
plays no further role in the ground state energy fluctuations.
The fluctuations therefore become sub-linear in
$U_0$, and eventually saturate to an interaction independent value.
Moreover, the results for the fluctuations become strongly geometry and
filling factor dependent \cite{Magic}.
We note that the observed saturation is in fact an artifact of the sharp
cut-off in the interaction range: with a longer-ranged interaction,
charge segregation cannot eliminate contributions due to the interaction
(although it may significantly reduce them), and the fluctuations
would no longer be bounded simply by kinetic energy considerations.

In Fig. \ref{exsrv} it can also be seen that for strong interactions, the
fluctuations are overestimated by $\Delta_2^{k_1}$ and $\Delta_2^{k_2}$, and
underestimated by $\Delta_2^{k_3}$. This can be understood within the picture
given above of charge density modulations. In this case, an occupied state
that is removed non-self-consistently will yield less energy than can
be gained when the system is allowed to reorganise, but the typical
size of this error saturates to an interaction independent value for
the same reason that the SCHF fluctuations do. If on the other hand
an unoccupied state is occupied non-self-consistently, it is not
possible to avoid contributions from the short-ranged part of the
potential (we do not consider strongly Anderson localised states at
very low filling), and the typical error increases indefinitely with
the interaction strength. As a result
$\Delta_2^{k_3}$ underestimates the fluctuations by an amount that saturates
to an interaction independent value, whereas fluctuations in the charging
energy predicted by $\Delta_2^{k_1}$ and $\Delta_2^{k_2}$ grow
with $U_0$ indefinitely; the errors made in employing the latter
approximations diverge with the interaction strength.

\begin{figure}[h]
\centerline{
\epsfysize 7.5cm
\epsffile{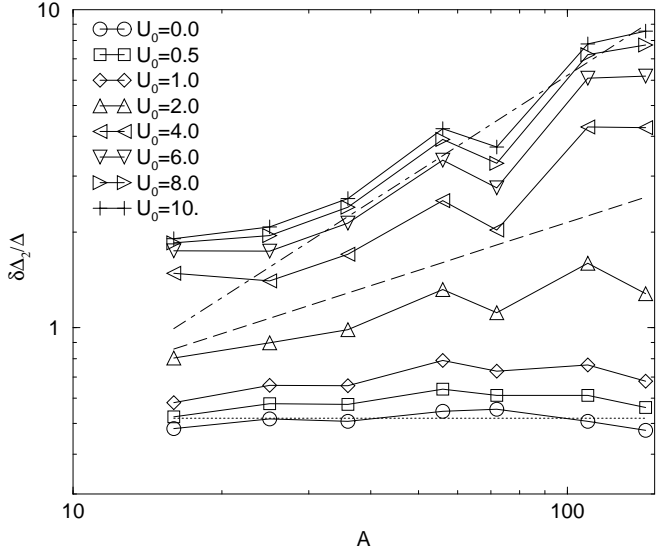}}
\caption{$\delta\Delta_2/\Delta$ against $A$, with
$W=4$ and $\nu\approx 1/4$, after 75 to 1000 disorder realisations.
The RMT results is shown as a dotted line, the dashed line is
proportional to $L$ and the dot-dashed line is proportional to $A$.
$r_s \approx U_0/2t$.}
\label{d2sra}
\end{figure}

Concentrating now on the fully self consistent results, the fluctuations in
the charging energy are plotted against the sample size in figure
\ref{d2sra}. As expected, for very weak interactions, the typical fluctuations
vanish like $1/A$, being dominated by kinetic energy fluctuations.
For stronger interactions, this dependence no longer holds: for $r_s\ll 1$
our results are in broad agreement with Ref.\cite{Blanter97}, but
do not agree with their suggestion that the typical fluctuations
remain proportional to $\Delta$ for $r_s >{\cal O}(1)$. This appears
to conflict with a simple single-parameter scaling argument \cite{Mirlin}.
The appearance of fluctuations that do not scale with $\Delta$ coincides with
the appearance of density modulations.
We stress that in this model there can be no physical connection between the
amplitude of the constant interaction and the amplitude of the fluctuations.

\begin{figure}[h]
\centerline{
\epsfysize 7.5cm
\epsffile{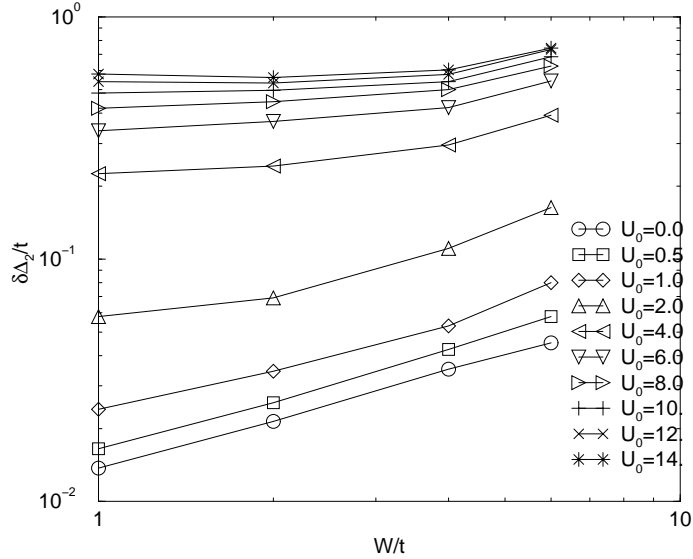}}
\caption{A typical result for $\delta\Delta_2/t$ against $W$,
averaged over 200 disorder realisations.
Here the sample is $11*10$, $N=28$.$r_s \approx 0.54 U_0/t$.}
\label{d2srw}
\end{figure}

For strong interactions the dominant disorder dependence appears to
develop only for $W\agt 4$, where it is consistent with the emergence
of a linear dependence to be expected from spatial rearrangements
in the disorder potential. An example is plotted in Fig. \ref{d2srw}.
We reiterate that we also find strong geometry and filling factor
dependences. It is extremely difficult to extract disorder scalings
in such small systems because $W/t$ is required to be fairly large
to generate diffusive motion, which in turn stretches the spectrum
in the tails. This can be seen at weak interaction, where one would
have hoped to see a disorder independent plateau in $\Delta$ (i.e 
$\delta\Delta_2$ at $U_0=0$).

To summarise then, we find
$\delta\Delta_2\sim 0.52 \Delta +a r_s \Delta + {\cal O}(r_s^2)$,
where $a$ is an undetermined constant or function
of disorder strength. We note that the disorder scaling is
not clear because of the residual dependence of $\Delta$ on $W$.

%
\subsection{Long Range Interactions}
%
We consider here the results for the Coulombic bare potential.

\begin{figure}[h]
\centerline{(a)
\epsfysize 5cm
\epsffile{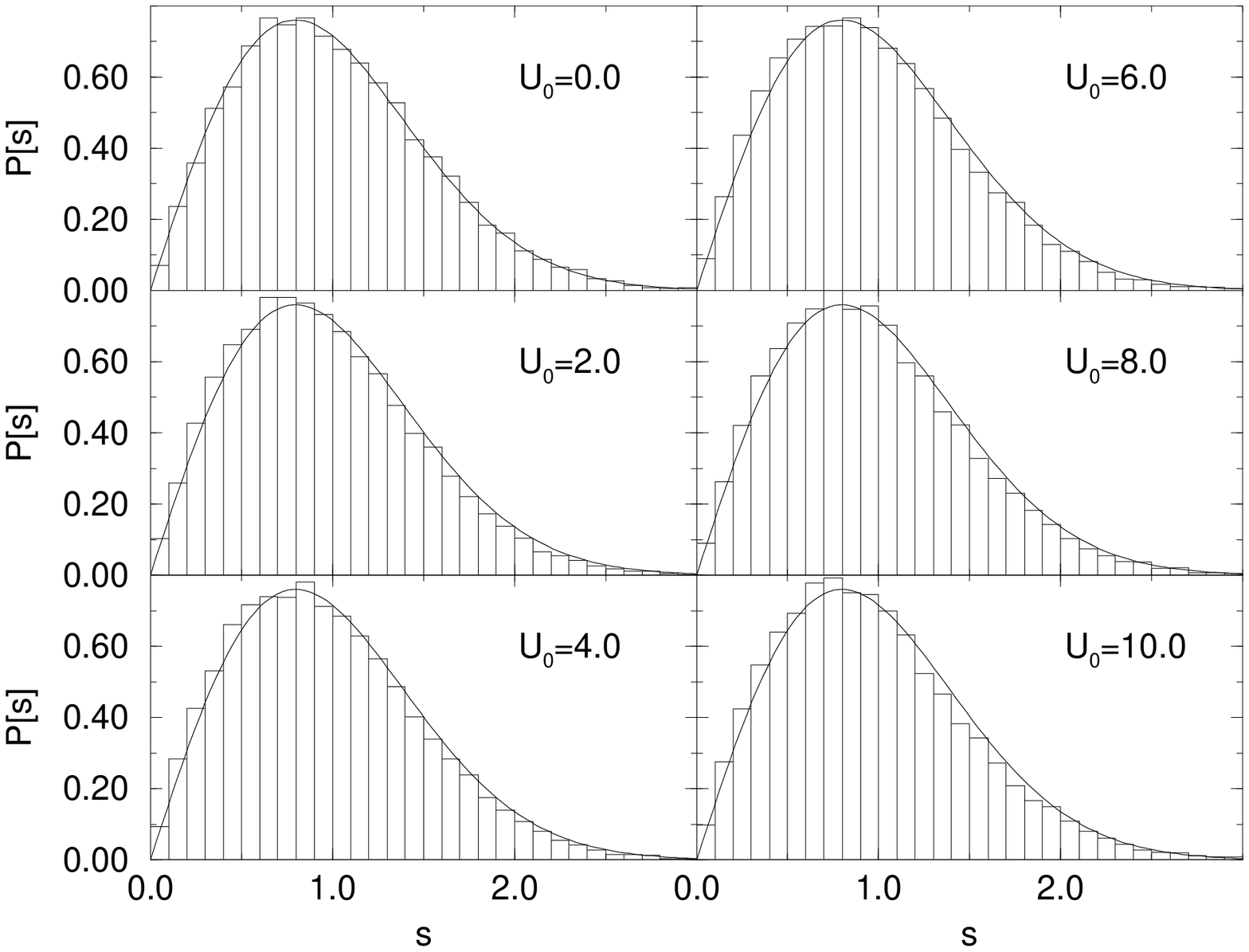}
\hspace{.5cm}(b)
\epsfysize 5cm
\epsffile{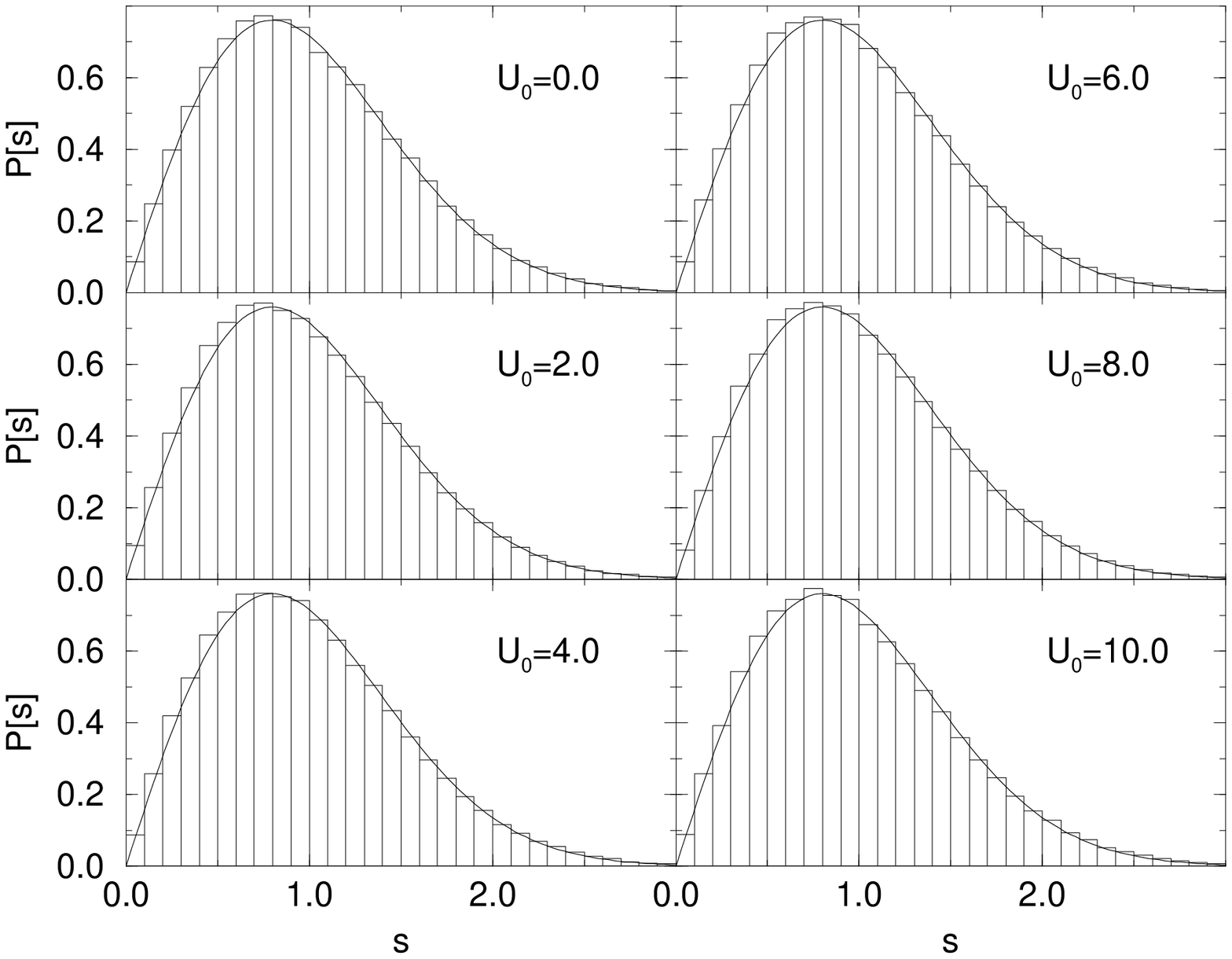}
}
\centerline{(c)
\epsfysize 5cm
\epsffile{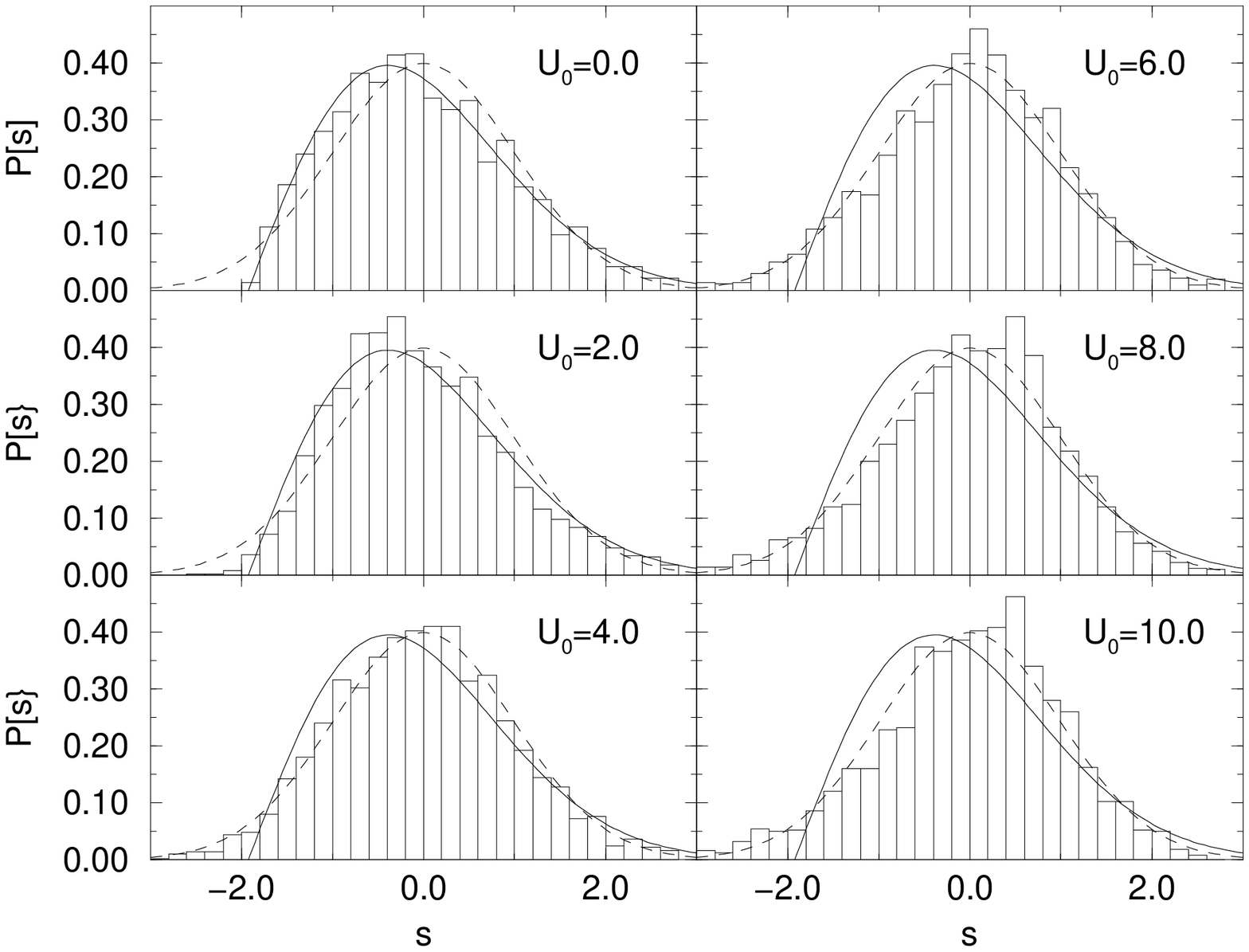}
}
\caption{SCHF level spacing distributions $P[s]$ for a) occupied states,
b) unoccupied states; $s=\Delta E/\langle\Delta E\rangle$, and c) the gap
$\Delta_2^{k_1}$, where
$s=(\Delta_2^{k_1} - \langle\Delta_2^{k_1}\rangle)/\delta\Delta_2^{k_1}$.
The solid lines show the WD distribution,
and in (c) the dashed line follows a Gaussian law. The samples were $8*9$
lattices with 14 electrons and Coulomb interactions; $W=2$.
$r_s \approx 0.56 U_0/t$. The statistics were
obtained from an ensemble of 2500 samples.}
\label{pslr}
\end{figure}

We first study the distributions of both the level spacings (\ref{nnspace})
and the gap (\ref{noSC}) of the SCHF spectrum at finite $r_s$, $g$.
In this case (\ref{noSC}) and (\ref{nnspace}) are calculated
in the self-consistent basis of $N$ particles.  In Fig. \ref{pslr} it is seen that
the normalised level spacings between occupied states obey statistics very
close to WD for all interaction strengths considered. Between occupied states
(Fig. \ref{pslr}a) there is a mild deviation towards Poisson statistics for the
strongest interaction strengths, indicative of a weak tendency towards
localisation. Between unoccupied states (Fig. \ref{pslr}b) the distribution is
even closer to WD.
On the other hand, the normalised gap distribution clearly tends towards a
more symmetric distribution that is approximately Gaussian.

\begin{figure}[h]
\centerline{
\epsfysize 7.5cm
\epsffile{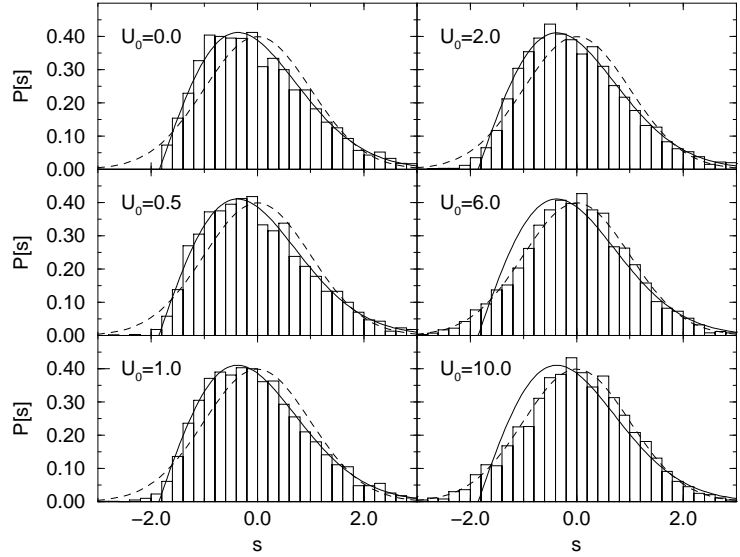}}
\caption{Distributions $P[s]$ where
$s=(\Delta_2-\langle\Delta_2\rangle)/\delta\Delta_2$
for various interaction strengths
obtained self consistently for the Coulomb interaction.
The solid line show the WD distribution, the dashed line shows the
Gaussian distribution. The samples were $7*8$ lattices with $N=15$, $W=4$,
and the statistics were obtained from an ensemble of 3000 samples.
$r_s \approx 0.56 U_0/t$.}
\label{pd2lr}
\end{figure}

We have also investigated the gap ($\Delta_2$) distribution obtained within
the fully self-consistent scheme, which we show in Fig. \ref{pd2lr}.
Again we find
that as $U_0$ is increased, the distribution evolves from a WD form to
a symmetric distribution similar to a Gaussian.

\begin{figure}[h]
\centerline{
\epsfysize 7.5cm
\epsffile{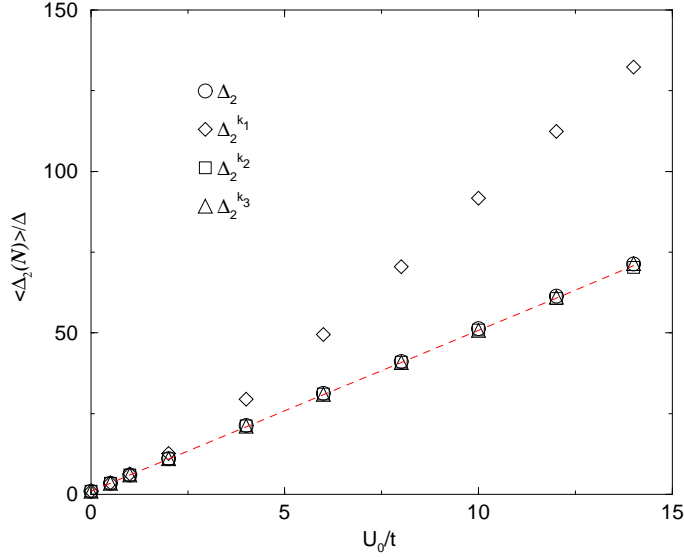}}
\caption{Typical results for the mean Coulomb gap under various approximation
schemes, averaged over 400 disorder realisations. Here, $W=4$, the lattice is
$11*10$ and $N=28$. $r_s \approx 0.54 U_0/t$.
The dashed line is the CI result for the mean.}
\label{lrexm}
\end{figure}

We shall concentrate first on the mean values obtained from these
distributions,
and will come to the variance later in the section. Figure \ref{lrexm}
shows a  comparison of $\langle\Delta_2\rangle$,
with the various approximations to it, plotted against $U_0$.
Whilst $\langle\Delta_2^{k_2}\rangle$ and
$\langle\Delta_2^{k_3}\rangle$ provide a good approximation to
$\langle\Delta_2\rangle$, we see a clear deviation of
$\langle\Delta_2^{k_1}\rangle$.
Results for the CI model, evaluated as described above, are
plotted for comparison. That the CI model is good in the mean indicates
that the single particle wavefunctions remain roughly uniformly
distributed over the dot for all $r_s$
considered.

\begin{figure}[h]
\centerline{
\epsfysize 7.5cm
\epsffile{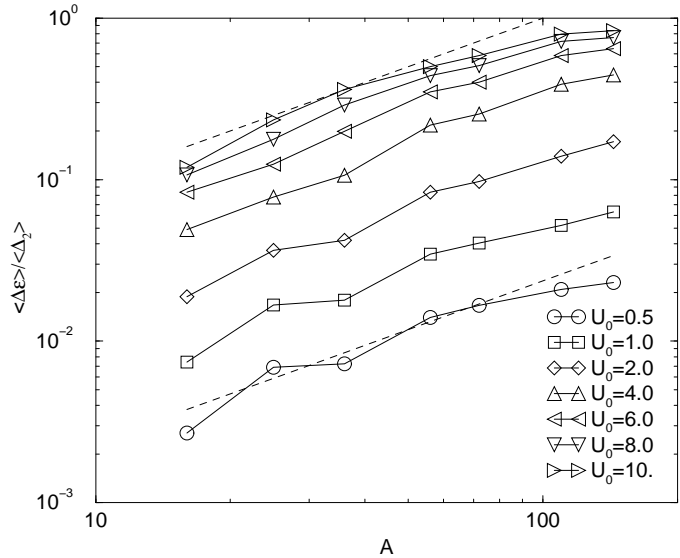}}
\caption{ $\langle\Delta\epsilon\rangle/\langle\Delta_2\rangle$
against the sample area $A$; $W=4$,
$\nu\approx 1/4$, and the results averaged over 300 to 1000 disorder
realisations, for a range of interaction strengths. The dashed lines show
$\langle\Delta\epsilon\rangle/\langle\Delta_2\rangle\propto A$ as guides
for the eye. $r_s \approx U_0/2t$.
}
\label{delra}
\end{figure}

In figure \ref{delra} we plot
$\langle\Delta\epsilon\rangle/\langle\Delta_2\rangle$ against
the sample area $A$, for an intermediate disorder strength ($W=4$).
Since we know from Fig.\ref{lrexm} that 
$\langle\Delta_2^{k_2}\rangle\approx\langle\Delta_2\rangle$, then
$\langle\Delta\epsilon\rangle$ is very close to $\langle\Delta_2^{k_1}\rangle-
\langle\Delta_2\rangle$, the total error made by applying Koopmans' theorem.
For $A\alt 50$, we find that for a fixed interaction strength
$\langle\Delta\epsilon\rangle\propto L$
($\langle\Delta\epsilon\rangle/\langle\Delta_2\rangle\propto A$).
For larger samples with $U_0\alt 2$ we find a weakening in the
dependence, but see no indication that it will vanish relative to $\Delta$.
The result that the deviations from Koopmans' approximation increase with
system size (when compared to $\Delta$), showing no sign of saturation,
is admittedly strange, and may be an artifact of the specific model
considered here. However, the result that Koopmans' approximation appears
to fail even as the system size tends towards the thermodynamic limit,
is in line with our findings for the short-ranged case \cite{Thermodynamic}.

\begin{figure}[h]
\centerline{
\epsfysize 7.5cm
\epsffile{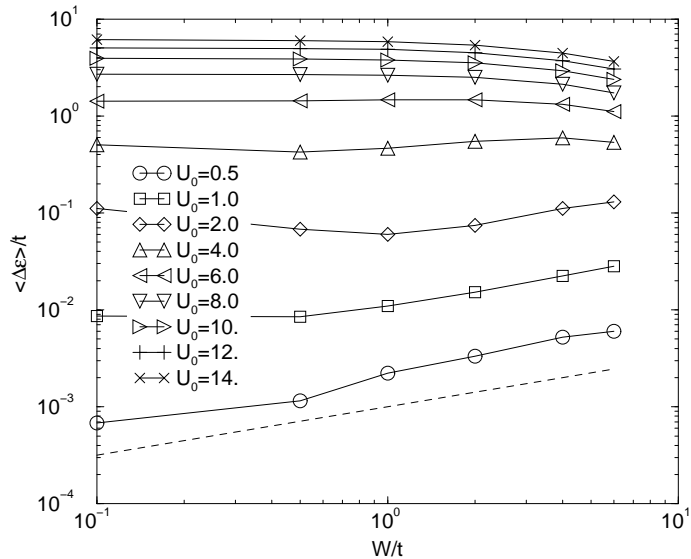}}
\caption{$\langle\Delta\epsilon\rangle/\Delta$ against disorder $W/t$
averaged over a ensembles of 300, $11*10$ samples with 28 particles.
$r_s \approx 0.54 U_0/t$.
The dashed line is proportional to $\protect\sqrt{W}$.}
\label{delrw}
\end{figure}

It is interesting to see how the error depends on disorder. 
In figure \ref{delrw} we plot $\langle\Delta\epsilon\rangle/t$ for a range of
disorder strengths: the disorder dependence as a function of interaction is
weak, but not simple. Similarly to the short-ranged case, the deviations
from Koopmans' approximation do not decrease for small disorder. This occurs
for the same reasons as for the short-ranged case. Here too the typical size
of the matrix elements $V_{ijkl}$ which drive the rearrangement scale
inversely with $g$. One thus expects that 
in this regime $\langle\Delta\epsilon\rangle$ should
increase with disorder, and this is indeed seen in the figure.
For $U_0\agt 4t$, $\Delta\epsilon$ decreases with disorder at sufficiently
large $W$, with evidence of a turning point ($d\Delta\epsilon/dW=0$) at
$U_0=4t$, $W\approx 4t$.

\begin{figure}[h]
\centerline{
\epsfysize 7.5cm
\epsffile{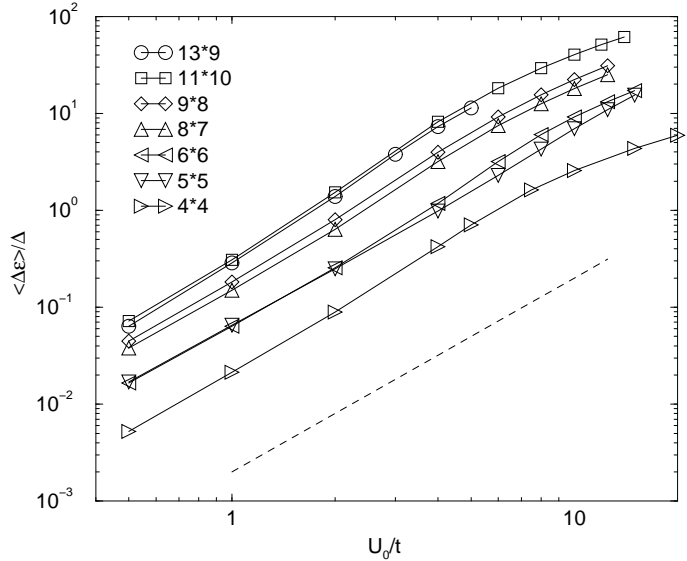}}
\caption{$\langle\Delta\epsilon\rangle/\Delta$ against $U_0/t$ after 300 to
1000 disorder realisations, for a range of sample sizes, all at approximately
quarter filling with $W=4$. $r_s \approx U_0/2t$.
The dashed line is proportional to $U_0^2$.}
\label{delru}
\end{figure}

In figure \ref{delru} we plot the interaction dependence of
$\langle\Delta\epsilon\rangle/\Delta$. We find that at small $U_0$,
$\langle\Delta\epsilon\rangle/\Delta\propto (U_0/t)^2$, with deviations
for larger $U_0$. This quadratic behaviour has the same origin as that of
the short-ranged case: the indication is that second order
perturbation theory is qualitatively good even for $r_s \sim 1$.

To summarise the results for the mean Coulomb gap,
we find that Koopmans' approximation (\ref{k1}) makes an
error in the mean charging energy which for small $r_s$ and $L$, and fixed
disorder scales like $\langle\Delta\epsilon\rangle\propto r_s^2 L$.
There is also evidence that for the larger sizes ($A\agt 50$), far
beyond that accessible by exact diagonalisation, that the size dependence
vanishes: $\langle\Delta\epsilon\rangle\propto r_s^2$.
In contrast to the naive expectation however, we find no
sign of this error vanishing relative to $\Delta$ in the thermodynamic
limit. This is due to the fundamental difference between occupied and
unoccupied SCHF levels already discussed in the short-ranged case.

\begin{figure}[h]
\centerline{
\epsfysize 7.5cm
\epsffile{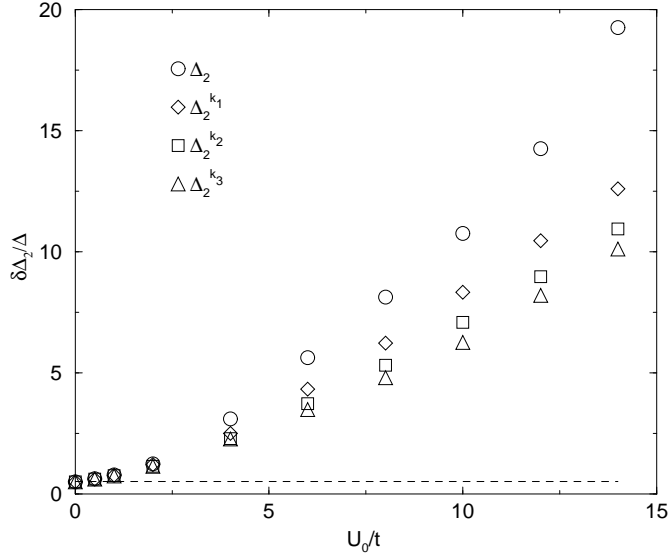}}
\caption{Typical results for $\delta\Delta_2/\Delta$ under various
approximation schemes, averaged over 300 disorder realisations. Here,
$W=4$, the lattice is $11*10$ and $N=28$, $r_s \approx 0.54 U_0/t$.
The Koopmans' approximants can be
seen to underestimate the fluctuations.}
\label{lrexv}
\end{figure}

\begin{figure}[h]
\centerline{
\epsfysize 7.5cm
\epsffile{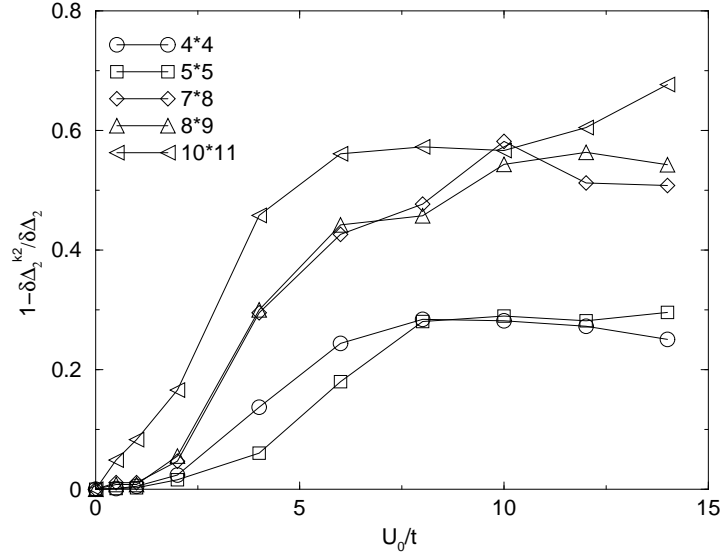}}
\caption{$\delta\Delta_2^{k_2}/\delta\Delta_2$ against $U_0$ for a
range of sample sizes, at approximately quarter filling:
$r_s \approx U_0/2t$. The statistics
are obtained from 300 to 1000 disorder realisations for each sample size.}
\label{dd2k}
\end{figure}

We now consider the fluctuations in $\Delta_2$. As an example of the
interaction dependence of these fluctuations in the various approximation
schemes, we plot the results for a fixed size in figure \ref{lrexv}.
It is seen that applying Koopmans' theorem in the forms
(\ref{k1}-\ref{k3}) results in considerably smaller fluctuations
than the fully self-consistent calculation. To quantify this error, we
plot $\delta\Delta_2^{k_2}/\delta\Delta_2$ in Fig.\ref{dd2k}, which shows
that the relative error initially increases with interaction strength, but
shows signs of saturating. The value of the saturation appears to increase
towards unity as the system size is increased.

\begin{figure}[h]
\centerline{
\epsfysize 7.5cm
\epsffile{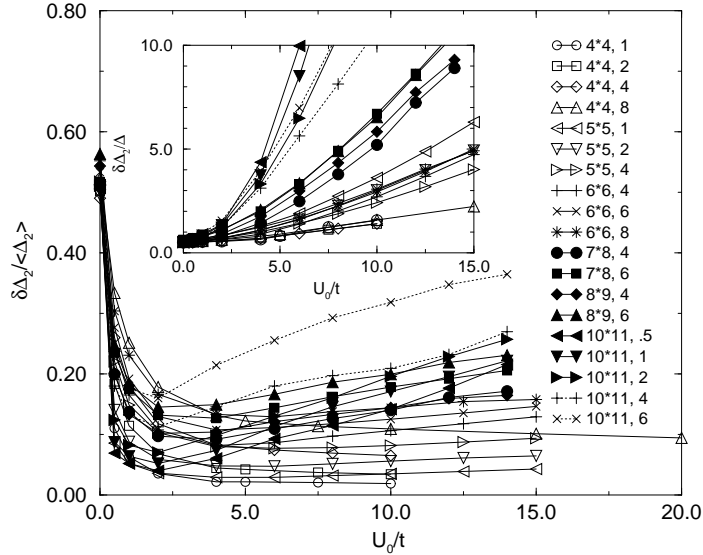}}
\caption{$\delta\Delta_2/\langle\Delta_2\rangle$ against $U_0/t$, after
300 to 1000 disorder realisations. The legend shows the sample size and $W$,
$\nu\approx 1/4$: $r_s \approx U_0/2t$.
The random matrix theory result at $U_0=0$ is 
$\delta\Delta_2/\langle\Delta_2\rangle\approx 0.52$. Inset:
$\delta\Delta_2/\Delta$ for the same data set.}
\label{dd2_d2}
\end{figure}

We now concentrate on the fluctuations of the fully self-consistent
peak spacing $\Delta_2$.
For comparison with Ref.\cite{Sivan96} it is useful to plot
$\delta \Delta_2/\langle\Delta_2\rangle$ against the interaction strength
for a range of sample sizes. This is done in figure \ref{dd2_d2}.
In the inset we plot $\delta \Delta_2/\Delta$ which shows that the
peak spacing fluctuations are not proportional to $\Delta$ for
$r_s L\sim{\cal O}(1)$ \cite{Coulomb}.
From Fig. \ref{dd2_d2} it can be seen that the curves
$\delta \Delta_2/\langle\Delta_2\rangle$
do not saturate to a constant as suggested in Ref.\cite{Sivan96}, although
to see this clearly one has to consider larger sample sizes than are
accessible by exact calculations.
The curve $\delta \Delta_2/\langle\Delta_2\rangle$ appears to take on the
approximate form of a constant term plus a linear term for
$r_s L\sim{\cal O}(1)$. The constant contribution identified by
Sivan {\it et el.} \cite{Sivan96}, is here, contrary to their claim,
non-universal (i.e. it is disorder dependent).

\begin{figure}[h]
\centerline{
\epsfysize 7.5cm
\epsffile{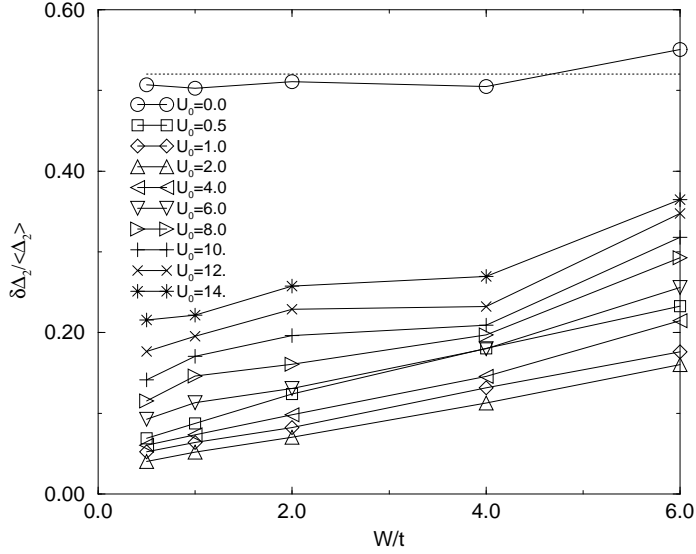}}
\caption{$\delta\Delta_2/\langle\Delta_2\rangle$ against $W/t$ averaged
over 300 $10*11$ samples with $N=28$. $r_s \approx 0.54 U_0/t$.
Results for other sample sizes were
similar. The CI + RMT result for $U_0 =0$ is plotted as a dotted line.}
\label{d2lrw}
\end{figure}

In figure \ref{d2lrw} we plot $\delta \Delta_2/\langle\Delta_2\rangle$ against
disorder for the $10\times 11$ lattice with a range of interaction strengths.
At $U_0=0$ it is seen
that for $W\alt 6$ the systems obeys WD statistics quite well. For the sample
size considered we find that in the regime $0.5\alt U_0\alt 6.0$
($0.25 \alt r_s \alt 3.0$) $\delta \Delta_2/\langle\Delta_2\rangle\propto W$,
and at stronger interactions this dependence weakens. The intermediate
dependence, $\delta \Delta_2/\langle\Delta_2\rangle\propto W$,
is consistent with the dependence
$\delta \Delta_2/\langle\Delta_2\rangle\propto 1/\sqrt{g}$
recently observed independently by Bonci and Berkovits \cite{Bonci}
for the Buminovich Stadium billiard. Analysis
of Fig. \ref{dd2_d2} leads to the conclusion that the quadratic contribution
(in $U_0$) to $\delta \Delta_2$ is
independent of disorder, which is consistent with Fig. \ref{d2lrw}.

\begin{figure}[h]
\centerline{
\epsfysize 7.5cm
\epsffile{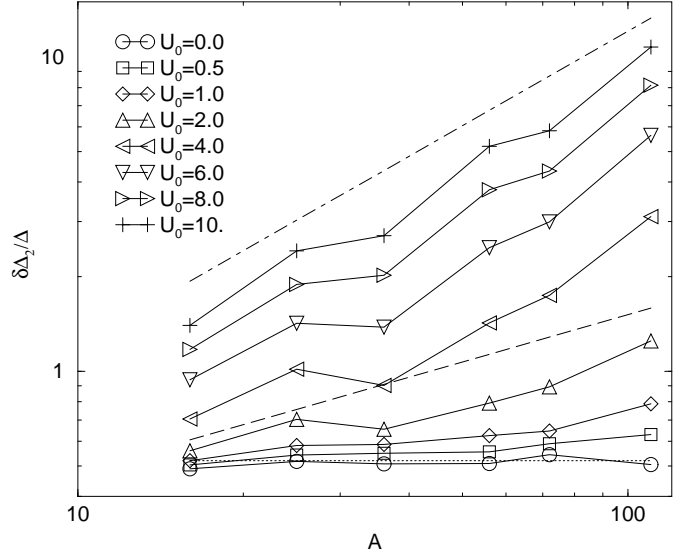}}
\caption{$\delta\Delta_2/\Delta$ against sample area $A$ averaged over
300 to 1000 disorder configurations. The dotted line shows the CI + RMT
result, the dashed line is proportional to $\protect\sqrt{A}$, and the
dot-dashed line is proportional to $A$. $r_s \approx U_0/2t$.}
\label{d2lra}
\end{figure}

To identify the system size scaling of the various contributions
we plot $\delta \Delta_2/\Delta$ against $A$ in figure \ref{d2lra}.
For $U_0=0$ the system obeys WD statistics and $\delta \Delta_2/\Delta$
is independent of size.
The regime over which the fluctuations are approximately proportional to
the mean charging energy (the constant contribution to
$\delta\Delta_2/\langle\Delta_2\rangle$ alluded to above, which
corresponds to a $\sqrt{A}$ dependence in the figure), depends on the system
size. As could be seen in Fig. \ref{dd2_d2} another term begins to
dominate the the fluctuations at larger $U_0$ which is quadratic in $U_0$,
this term increases
more rapidly with the system size, and so dominates at lower $U_0$
in larger systems. Over the range of sizes considered,
this term appears to scale like $L^2$.
The cross-over in dominance therefore occurs at $U_0\sim 1/L$
for fixed disorder strength; clearly the quadratic term will dominate in
large samples. The increase in $\delta\Delta_2/\Delta$ with system size
appears to be an artifact of using the unscreened Coulomb interaction
in the Hamiltonian.

Summarising the results presented in figures \ref{dd2_d2} to \ref{d2lra},
and the above discussion, we find for $r_s L \agt {\cal O}(1)$ an approximate
form: $\delta \Delta_2\sim.52 \Delta + a\langle\Delta_2\rangle/\sqrt{g}
+b r_s^2 $ where $a,b$ are constants.
One would normally expect the fluctuations to be linear in the interaction
strength, (i.e. $b=0$). A possible source for such a quadratic interaction
dependence in the typical fluctuations is the development of correlations
that grow like $r_s^2$ in products of eight wavefunctions. Elsewhere
\cite{Walker} we present evidence for increased fluctuations in the ground
state density in this regime, as compared to a non-interacting system.
It is not yet clear whether this result is an artifact of the SCHF
approximation (which has also very recently been observed in $1d$ systems
\cite{Richter} using a similar approximation scheme), or a genuine physical
effect.

Finally, it is worth noting that since the exchange interaction is not
correctly screened, that errors in the SCHF scheme might be expected to
diverge with respect to $\Delta$ as the system size is increased.
We can neither confirm nor counter this
argument, but have verified that for $r_s \alt 5$ the fluctuations in
the exchange contribution are smaller than those of the direct contribution.
%
\section{Summary}
%
We have investigated the addition spectra of disordered quantum dots
employing an effective single particle approximation, both
using a fully self consistent analysis, and by invoking Koopmans'
theorem. We were able to consider system sizes with up to 144 sites, and
37 particles, compared to the latest exact calculations on
samples with 24 sites and 6 particles \cite{Sivan96}. The larger sample
size also allows us to consider smaller values of $g$ than exact calculations
whilst retaining an ergodic non-interacting limit, and therefore approaches
the experimental parameters more closely.
The inclusion of spin in a consistent manner is left for a future project.

Our SCHF results for the typical fluctuations of the peak spacings
for particles possessing short-ranged bare interactions are entirely
different from the results for long-ranged bare interactions.
In the short-ranged case we find the same scaling as Ref.\cite{Blanter97}
for very weak interactions ($r_s\ll 1$), but for $r_s\agt 1$
deviations from this behaviour
become significant and coincide with the appearance of interaction
induced density modulations. We find no size dependence in the onset
of these effects. We find that strong filling factor and geometry
dependences arise due to these density fluctuations, and therefore do not
expect that the disorder ensemble statistics can be mapped to statistics
over the ensemble of filling factors: ergodicity is lost.
We suggest that employing a short ranged bare interaction in a
self-consistent scheme is not an appropriate model for the quantum dots
of Refs.\cite{Sivan96,Simmel,Marcus97,Marcus96,Marcus98,Patel,Chang95}
for which $r_s >1$, but may be a useful model for
dot geometries sandwiched between very close metallic
gates which provide a good external source of screening \cite{Ando,Gates}.
In this respect
we identify some experiments on the addition spectrum which possess a
metallic source (heavily doped $n^+$ GaAs) and drain (Cr/Au), at
separations of the order of the average inter-particle spacing in the dot
\cite{Ashoori92,Ashoori97}.

In the Coulomb case the SCHF approximation to $\Delta_2$ yields typical
fluctuations that do not scale with $\Delta$ for $r_s\agt 1/L$.
In Ref.\cite{Sivan96} it is claimed that $\delta\Delta_2$ is universally
proportional to $\langle\Delta_2\rangle$ for strong interactions (but
still far from the accepted Wigner Crystal transition point). In contrast,
we find, in addition to the small interaction independent contribution, a
contribution to $\delta\Delta_2$ that is proportional to
$\langle\Delta_2\rangle/\sqrt{g}$ (i.e. non-universal), and a further
contribution that scales like $r_s^2$, which is independent of disorder,
and appears to be due to the development of charge density modulations
\cite{Walker}. The latter is not detectable in the small
systems examined numerically in \cite{Sivan96}, and so our results
are not numerically inconsistent with exact calculations \cite{Sivan96}.
Whilst we do not include spin, the observed decrease in the fluctuations
with $g$ is consistent with the experimental indications
\cite{Sivan96,Marcus97} that in cleaner samples the fluctuations are smaller.

We show that a direct application of Koopmans' theorem overestimates
$\langle\Delta_2\rangle$. This overestimate, a manifestation of the breakdown
of Koopmans' approximation, does not vanish on the scale of $\Delta$ in the
thermodynamic limit. The error seems to scale differently with sample size
for sample areas above or below $A\approx 50$. In the nearest neighbour case,
with $A\agt 50$, this error scales with $\Delta$, but in smaller systems,
accessible by exact methods, it is independent of system size. In the
Coulomb case the error grows with the system size as $L$ for $A\alt 50$.
For larger sizes the error appears to tend towards a $1/L$ scaling, i.e.
in proportion with the charging energy, and therefore diverges with respect
to the mean effective single-particle level spacing. This result for the
Coulomb interaction case appears to be non-physical, and may be an artifact
of the model considered. However, the result that Koopmans' theorem is not
recovered in the thermodynamic limit also occurs in the short-ranged
interaction case. In both cases
we find that initially this error grows in proportion to $U_0^2$,
to be expected since the lowest order contribution is second order,
but for strong interactions it grows more slowly in $U_0$, and that
the disorder dependence of this error is weak and non-monotonic.
We identify the source of the error $\langle\Delta\epsilon\rangle$ to
be the fundamental difference between occupied and unoccupied states that
is inherent in the SCHF approximation. We introduce two improved
applications of Koopmans' theorem, $\langle\Delta_2^{k_2}\rangle$,
$\langle\Delta_2^{k_3}\rangle$, which provide a good approximation to 
$\langle\Delta_2\rangle$, but not to $\langle\delta\Delta_2\rangle$.

Whilst preparing the manuscript, two related works appeared that confirm
some of the points discussed above \cite{Bonci,Levit}.

In both cases, fluctuations in the ground state density develop with $r_s$
\cite{Walker}, and have significant effects of the addition spectrum
statistics. It remains to be seen whether these density modulations are an
artifact of the SCHF approximation (i.e. due the neglect of dynamical
correlations), or in fact interesting results on the continuous transition to a
Wigner-type solid in disordered samples with short- and long-ranged
bare interactions.
\acknowledgments
We acknowledge discussions with H. Orland and F. von Oppen in the early
stages of this project, as well as with Ya. Blanter, S. Levit, A. Mirlin
and D. Orgad.
We acknowledge support from the EU TMR fellowship ERBFMICT961202,
the German-Israeli Foundation, the U.S.-Israel Binational-Science
Foundation and the Minerva Foundation. One of us (YG) would also
like to acknowledge support from an EPSRC senior professorial fellowship,
grant number GR/L67103.
Much of the numerical work was performed using IDRIS facilities.

%
%

\begin{thebibliography}{99}
\bibitem{Koopmans}T. Koopmans Physica {\bf 1}, 104 (1934).
\bibitem{AGD}A. A. Abrikosov, L. P. Gorkov and I. E. Dzyaloshinski
Methods of Quantum Field Theory in Statistical Physics,
(Dover; New York, 1963).
\bibitem{Sivan96} U. Sivan, R. Berkovits, Y. Aloni, O. Prus, A. Auerbach
and G. Ben-Yoseph, Phys. Rev. Lett. {\bf 77}, 1123 (1996).
\bibitem{Simmel}F. Simmel, T. Heinzel and D. A. Wharam,
Europhys. Lett. {\bf 38}, 123 (1997).
\bibitem{Marcus97} S.R. Patel, S. M. Cronenwett, D. R. Stewart, A. G. Huibers,
C. M. Marcus, C. I. Dur\" oz, J. S. Harris Jr., K. Campman and A. C. Gossard,
Phys. Rev. Lett. {\bf 80}, 4522 (1998).
\bibitem{Ashoori92}R. C. Ashoori, H. L. Stormer, J. S. Weiner, L. N. Pfeiffer,
S. J. Pearson, K. W. Baldwin and K. W. West, Phys. Rev. Lett.{\bf 68}, 3088
(1992).
\bibitem{Ashoori97}N. B. Zhitnev, R. C. Ashoori, L. N. Pfeiffer and K. W. West,
cond-mat 9703241.
\bibitem{Efetov83}K. B. Efetov, Adv. Phys. {\bf 32}, 53 (1983).
\bibitem{Mehta}M. L. Mehta, Random Matrices 2nd ed.,
(Academic; New York, 1991).
\bibitem{Kravstov}V. E. Kravstov and A. D. Mirlin, JETP Lett. {\bf 60}, 645
(1994).
\bibitem{Marcus96}J. A. Folk, S. R. Patel, S. F. Godijn, A. G. Huibers,
S. M. Cronenwett, C. M. Marcus, K. Campman and A. C. Gossard,
Phys. Rev. Lett. {\bf 76}, 1699 (1996).
\bibitem{Marcus98}S. M. Cronenwett, S. R. Patel, C. M. Marcus,
K. Campman and A. C. Gossard, Phys. Rev. Lett. {\bf 79}, 2312 (1997).
\bibitem{Patel}S. R. Patel, D. R. Stewart, C. M. Marcus, M. G\" ok\c ceda\v g,
Y. Alhassid, A. D. Stone,  C. I. Dur\" oz and J. S. Harris Jr.,
Phys. Rev. Lett. {\bf 81}, 5900 (1998).
\bibitem{Chang95}A. M. Chang, H. U. Baranger, L. N. Pfeiffer, K. W. West and
T. Y. Chang, Phys. Rev. Lett. {\bf 76}, 1695 (1996).
\bibitem{Berkovits98}R. Berkovits and U. Sivan, Europhys. Lett. {\bf 41},
653 (1998).
\bibitem{Blanter97} Ya. M. Blanter, A. D. Mirlin and B. A. Muzykantskii,
Phys. Rev. Lett. {\bf 78}, 2449 (1997).
\bibitem{Stopa}M. Stopa, Physica {\bf B251}, 228, (1998).
\bibitem{Vallejos98} R. O. Vallejos,, C. H. Lewenkopf, E. R. Mucciolo,
cond-mat/9802124.
\bibitem{Poilblanc}G. Bouzerar and  D. Poilblanc, Phys. Rev. B {\bf 52},
10772 (1995); J. Phys. I France, {\bf 7}, 877 (1997).
\bibitem{BerkAlt}R. Berkovits and B. L. Altshuler Phys. Rev. {\bf B55}, 5297
(1997).
\bibitem{Distance}This shift can be quantified
by $1-|\{\langle\Psi^N|\otimes\langle\psi^N_{N+1}|\}|\Psi^{N+1}\rangle|^2$,
where $|\Psi^m\rangle$ is the ground state SCHF Slater determinant of
$m$ particles,
$|\psi^m_l\rangle$ is the $l$th single particle state of the $m$ particle
SCHF spectrum, and $\otimes$ is the antisymmetrised tensor product.
\bibitem{Ceperley}B. Tantar and D. Ceperley, Phys. Rev. {\bf B39}, 5005 (1989);
S. T. Chui and K. Esfarjani, Europhys. Lett. {\bf 14}, 361 (1991);
S. T. Chui and B. Tantar, Phys. Rev. Lett.{\bf 74}, 458 (1992).
\bibitem{Blanter}A. Kamenev and Y. Gefen (unpublished);
Ya. M. Blanter Phys. Rev {\bf B54}, 12807 (1996).
\bibitem{Walker}P. N. Walker, Y. Gefen and G. Montambaux, cond-mat/9902099.
\bibitem{Thouless}D. J. Thouless, The Quantum Mechanics of Many Body
Systems 2nd ed., (Academic; London, 1972).
\bibitem{Magic}This is particularly spectacular at certain special
filling factors \cite{Walker}.
\bibitem{Mirlin}A. D. Mirlin (private communication).
\bibitem{Thermodynamic}These results
cannot be directly extrapolated to the thermodynamic limit,
because at fixed disorder one would arrive at states that are localised
on a length scale short compared to the system size in the
limit $r_s \to 0$. Here however, the non-interacting wavefunctions are
extended over the entire sample.
\bibitem{Coulomb}The result that interactions become
important at increasingly low $r_s$ as $L$ is increased, rather than
a size independent density (e.g. $r_s \sim{\cal O}(1)$) is an artifact of
the long range of the bare interaction: The relevant interaction matrix
elements grow with $L$ relative to $\Delta$. It is well known that the Coulomb
interaction causes divergences (for $L\to\infty$), and is therefore usually
screened explicitly. Here however, the (static) screening is generated
self consistently. The results suggest that the screening requires
increasing rearrangement as $L$ is increased.
\bibitem{Bonci}L. Bonci and R. Berkovits cond-mat/9901332.
\bibitem{Richter}A. Cohen, K. Richter and R. Berkovits, Phys. Rev.
{\bf B57}, 6223 (1998).
\bibitem{Ando}T. Ando, A. B. Fowler and F. Stern, Rev. Mod. Phys.
{\bf 54}, 437 (1982).
\bibitem{Gates}
In the case of one close metallic gate, the interaction is dipolar ($1/r^3$)
at distances greater than the dot to gate separation, and in the case of
two close gates (above and below the dot) the the long range interactions
are exponentially small. The nearest neighbour interaction can be considered as
a model for such potentials.
\bibitem{Levit}S. Levit and D. Orgad, cond-mat/9901298.
\end{thebibliography}
\end{document}